\documentclass[useAMS,usenatbib]{mn2e}

\usepackage[T1]{fontenc}
\usepackage{pslatex}
\usepackage{amsmath}
\usepackage{amsfonts}
\usepackage{color}
\usepackage{url}
\usepackage{graphicx}
\usepackage{subfigure}
\usepackage{amssymb}
\usepackage{textcomp}
\usepackage{soul}
\usepackage{booktabs}
\usepackage{array}
\usepackage[draft]{hyperref}
 \graphicspath{{images/}}

{\newif\ifnotend
\notendtrue
\def\veclist{ABCDEFGHIJKLMNOPQRSTUVWXYZabcdefghijklmnopqrstuvwxyz.}
\def\top#1#2.{#1}
\def\tail#1#2.{#2.}
\loop\expandafter\xdef\csname v\expandafter\top\veclist\endcsname%
{{\noexpand\bf\expandafter\top\veclist}}
\edef\veclist{\expandafter\tail\veclist}
\if\veclist.\notendfalse\fi\ifnotend\repeat}

\def\pa{\partial}

\mathchardef\mhyphen="2D

\title[An explanation for the CMZ asymmetry]{A theoretical explanation for the Central Molecular Zone asymmetry}
\author[Sormani, Tre{\ss}, Ridley, Glover, Klessen, Binney, Magorrian, Smith]{Mattia C. Sormani$^{1}$, Robin G. Tre{\ss}$^1$, Matthew Ridley$^2$,  Simon C.O. Glover$^1$, \newauthor 
Ralf S. Klessen$^{1,3}$, James Binney$^2$, John Magorrian$^2$ and Rowan Smith$^4$\\
$^1$Universit\"{a}t Heidelberg, Zentrum f\"{u}r Astronomie, Institut f\"{u}r theoretische Astrophysik, Albert-Ueberle-Str. 2, 69120 Heidelberg, Germany \\
$^2$Rudolf Peierls Centre for Theoretical Physics, 1 Keble Road, Oxford OX1 3NP, UK\\
$^3$Universit\"at Heidelberg, Interdiszipli\"ares Zentrum f\"ur Wissenschaftliches Rechnen, Im Neuenheimer Feld 205, 69120 Heidelberg, Germany \\
$^4$Jodrell Bank Centre for Astrophysics, School of Physics and Astronomy, University of Manchester, Oxford Road, Manchester M13 9PL, UK
}

\begin{document}
\hyphenation{kruijs-sen}

\date{}

\def\p{\partial}
\def\Omegap{\Omega_{\rm p}}

\newcommand{\di}{\mathrm{d}}
\newcommand{\bfx}{\mathbf{x}}
\newcommand{\bfe}{\mathbf{e}}
\newcommand{\vlos}{\mathrm{v}_{\rm los}}
\newcommand{\Tspin}{T_{\rm s}}
\newcommand{\Tb}{T_{\rm b}}
\newcommand{\degree}{\ensuremath{^\circ}}
\newcommand{\Th}{T_{\rm h}}
\newcommand{\Tc}{T_{\rm c}}
\newcommand{\bfr}{\mathbf{r}}
\newcommand{\bfv}{\mathbf{v}}
\newcommand{\pc}{\,{\rm pc}}
\newcommand{\kpc}{\,{\rm kpc}}
\newcommand{\Myr}{\,{\rm Myr}}
\newcommand{\Gyr}{\,{\rm Gyr}}
\newcommand{\kms}{\,{\rm km\, s^{-1}}}
\newcommand{\de}[2]{\frac{\partial #1}{\partial {#2}}}
\newcommand{\cs}{c_{\rm s}}
\newcommand{\rb}{r_{\rm b}}
\newcommand{\rqu}{r_{\rm q}}
\newcommand{\nuP}{\nu_{\rm P}}
\newcommand{\thetaobs}{\theta_{\rm obs}}
\newcommand{\hatn}{\hat{\textbf{n}}}
\newcommand{\hatx}{\hat{\textbf{x}}}
\newcommand{\haty}{\hat{\textbf{y}}}
\newcommand{\hatz}{\hat{\textbf{z}}}
\newcommand{\hatX}{\hat{\textbf{X}}}
\newcommand{\hatY}{\hat{\textbf{Y}}}
\newcommand{\hatZ}{\hat{\textbf{Z}}}
\newcommand{\hatN}{\hat{\textbf{N}}}

\maketitle

\begin{abstract}
It has been known for more than thirty years that the distribution of molecular gas in the innermost 300 parsecs of the Milky Way, the Central Molecular Zone, is strongly asymmetric. Indeed, approximately three quarters of molecular emission comes from positive longitudes, and only one quarter from negative longitudes. However, despite much theoretical effort, the origin of this asymmetry has remained a mystery. Here we show that the asymmetry can be neatly explained by unsteady flow of gas in a barred potential. We use high-resolution 3D hydrodynamical simulations coupled to a state-of-the-art chemical network. Despite the initial conditions and the bar potential being point-symmetric with respect to the Galactic Centre, asymmetries develop spontaneously due to the combination of a hydrodynamical instability known as the ``wiggle instability'' and the thermal instability. The observed asymmetry must be transient: observations made tens of megayears in the past or in the future would often show an asymmetry in the opposite sense. Fluctuations of amplitude comparable to the observed asymmetry occur for a large fraction of the time in our simulations, and suggest that the present is not an exceptional moment in the life of our Galaxy.
\end{abstract}


\section{Introduction}
\label{sec:intro}

It is now well established that the Milky Way (MW) is a barred galaxy, both from photometric \citep[e.g.][]{Blitz1991,Dwek1995,Binney1997,WeggGerhard2013} and dynamical evidence \citep[e.g.][]{MulderLiem1986,Binney++1991,EnglmaierGerhard1999,Bissantz++2003,RFC2008,SBM2015a,SBM2015c,Li++2016}; see also the reviews by \cite{FuxBarReview} and \cite{BlandHawthornGerhard2016}.

In much the same way that the gas flow in an axisymmetric disc galaxy can be understood considering circular orbits, the gas flow in a barred potential can be understood considering $x_1$ and $x_2$ orbits \citep[][]{Binney++1991}. $x_1$ orbits are closed orbits elongated parallel to the bar major axis. Gas in the outer parts of the bar follows these orbits while drifting slowly inwards along a sequence of such orbits. $x_1$ orbits become more and more elongated as the centre is approached, until the ``cusped orbit'' is reached. Interior to the cusped orbit, the orbits of the $x_1$ family become self-intersecting. At this point, the gas must transfer to $x_2$ orbits, which are elongated perpendicular to the bar major axis. The transition happens through the mediation of large-scale shocks, which correspond to observed dust lanes in external barred galaxies \citep{Athan92b}. The shocked gas plunges from the cusps of the cusped $x_1$ orbit towards the centre, until at some smaller radius it piles up and organises into a mildly elliptical disc/ring-like structure where gas follows $x_2$ orbits. This schematic picture is refined by hydrodynamical models that can properly take into account the effects of gas pressure neglected in the ballistic approximation (see for example \citealt{SBM2015a} for a detailed comparison of closed orbits vs a simple hydrodynamical simulation).

This framework can account for many prominent observational features in the region $| l | < 30 \degree$ of our Galaxy. For example, the $3\mhyphen\kpc$ arm, its far-side counterpart and the $135\mhyphen\kms$ Arm can be explained as spiral arms extending out from the ends of the bar, the Connecting Arm is usually identified with the near-side dust lane of the MW bar, and the observed high-velocity peaks observed in the terminal velocity curve of H{\sc I} and CO can be explained by strong non-circular motions created by the bar, i.e. by gas on the innermost populated $x_1$ orbits \citep[e.g.][]{Fux1999,EnglmaierGerhard1999,Bissantz++2003,RFC2008,SBM2015a,SBM2015c,Li++2016}. Simulations of gas flow in barred galaxies can also be used to constrain the properties of the MW and of external galaxies. For example, \cite{SBM2015c} used them to constrain the MW bar pattern speed, length and strength, and \cite{Fragkoudi+2017} used them to constrain the dark matter content of NGC 1291.

According to this framework, the Central Molecular Zone (CMZ), the innermost few hundred parsecs of the MW, should be identified with the mildly-elliptical gaseous disc/ring like structure made of gas flowing on $x_2$ orbits. In this context, \cite{Ridley+2017} studied the dynamics of gas in the CMZ.  Using simple isothermal simulations in an externally imposed barred potential they constructed a coherent picture of the gas flow, and related connected structures observed in $(l,b,v)$ data cubes of molecular tracers to nuclear spiral arms and other features that spontaneously appeared in the simulations. 

While several features present in $(l,b,v)$ data cubes of molecular emission were reproduced by the \cite{Ridley+2017} model, two key features remained unexplained:
\begin{enumerate}
\item Why is the CMZ asymmetric? It is well known that three quarters of the $^{13}$CO and CS emission comes from positive longitudes, and only one quarter from negative longitudes \citep{Bally1988}. The line of sight velocity distribution is also similarly asymmetric. Perspective effects cannot account for this \citep{JenkinsBinney94}. 

Based on the theoretical framework described above there seems to be no reason not to expect a calm, symmetric $x_2$ disc/ring. However, recently \cite{KimKimKim2014} and \cite{SSSTK} have formally shown that large-scale galactic shocks such as those developed as the gas transits from $x_1$ to $x_2$ orbits are unstable due to a hydrodynamical instability known as the ``wiggle instability'' \citep{WadaKoda2004} - for which a more appropriate name is perhaps Periodic Shock Instability (PSI, \citealt{SSSTK}). On account of this instability, unsteady flow is present in the simulations of \cite{SBM2015a} and \cite{Ridley+2017}. These authors conjectured that a possible explanation for the asymmetry is large-scale unsteady flow deriving from this instability. They argued that unsteady flow would cause unsteady conversion from atomic to molecular gas. However, their simulations did not keep track of the chemistry of the ISM, thus this remained a promising but speculative conjecture.

\item What are the ``vertical features'' observed in $(l,v)$ diagrams of molecular tracers? Near the Galactic Centre ($| l | \lesssim 5\degree$) there are several features that span a wide velocity range ($\sim 100 \kms$ or more) while being confined to a narrow longitude range (e.g. \citealt{Liszt2006}, see also \citealt{Fukui+2006,Suzuki+2015}). Bania Clump 2 is the prototypical example of such vertical features \citep[][]{StarkBania1986}. The velocity dispersions observed in these clouds are too large to be attributed to small-scale turbulence alone, and some large-scale galactic dynamical process must be involved. \cite{SBM2015c} and \cite{Ridley+2017} \citep[see also][]{Fux1999} proposed that most vertical features originate from different portions of the dust lanes of the MW. However, while the projections of the bar shocks (i.e., the dust lanes) in their models were in the right positions in the $(l,v)$ plane, they were not prominent enough. These authors argued that the projections of the bar shocks should be much more prominent in CO $(l,v)$ diagrams than their simple simulations suggested, thanks to efficient conversion of gas from atomic to molecular form at the shocks. However, they were unable to confirm this since their simulations did not keep track of the chemistry of the ISM, and this also remained merely at the stage of a promising conjecture.
\end{enumerate}
Validation of the conjectures mentioned in items (i) and (ii) requires high-resolution simulations that can keep track of the chemical composition of the ISM. 

In this paper we present high-resolution 3D simulations that include a state-of-the-art time-dependent chemical network, at the level of sophistication of those used in studies of individual molecular clouds \citep[e.g.][]{Bertram+2015}, but applied to the whole Galactic bar region. We use exactly the same potential as in \cite{Ridley+2017}. This allows for a direct comparison so we can isolate the effects of the new physical ingredients included here. Self-gravity of the gas is not included in either work, therefore there are only two differences between \cite{Ridley+2017} and the present work: i) the inclusion of the chemistry and of heating and cooling processes. These significantly affect the dynamics, since they make the gas subject to the thermal instability of \cite{Field1965}, providing a method for creating density inhomogeneities that was absent in the isothermal simulations of \cite{Ridley+2017}; ii) the simulations in this work are three-dimensional, while those in \cite{Ridley+2017} were two-dimensional.

In this paper we focus on the CMZ asymmetry and postpone the discussion of vertical features to future work. This paper is structured as follows. In Section \ref{sec:data} we briefly review the observational evidence for the asymmetry. In Section \ref{sec:methods} we outline our numerical methods. In Section \ref{sec:results} we discuss the dynamics of the gas flow. In Section \ref{sec:discuss} we discuss the implications of our models for the asymmetry of the CMZ and for the central regions of external galaxies. We also discuss the limitations of our models. Finally, in Section \ref{sec:conc} we summarise our conclusions.

\section{Data}
\label{sec:data}

\begin{figure}
\includegraphics[width=0.48\textwidth]{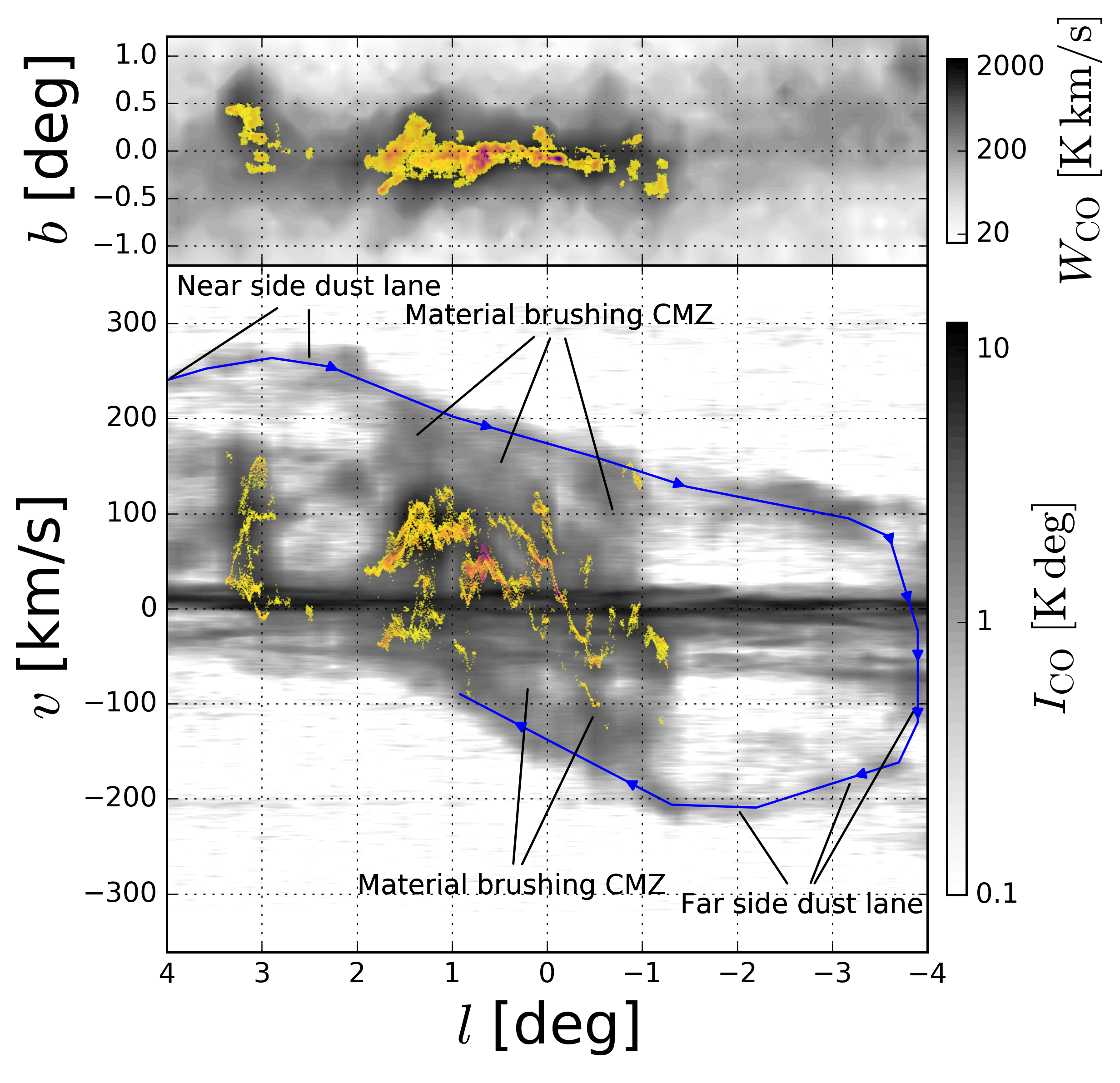}
\caption{NH$_3$ $(1-1)$ data from \protect\cite{Longmore2017} (yellow colorscale), superimposed on CO $(J=1\to0)$ emission from \protect\cite{Bitran+1997} (gray colorscale). Top panel: the ($l,b$) distribution integrated over $v$. Bottom panel: the position-velocity distribution $(l,v)$ integrated over $| b | \leq 2\degree$. Schematic lines and annotations are according to the interpretation discussed later in the text. Each NH$_3$ point represents a spectral component as determined by the \textsc{scouse} algorithm, coloured by brightness temperature. These data comes from the H$_2$O Southern Galactic Plane Survey (HOPS; \citealt{Walsh+2011}, \citealt{Purcell+2012}). They have a spatial resolution of $\sim~60\,\rm~arcsec$ and a spectral resolution of 2$\,\mathrm{km}\,\mathrm{s}^{-1}$, and have been fit using the Semi-automated multi-COmponent Universal Spectral-line fitting Engine (\textsc{scouse}, \citealt{Henshaw+16a}, \url{https://github.com/jdhenshaw/SCOUSE}).}
\label{fig:henshaw}
\end{figure}

Figure \ref{fig:henshaw} shows NH$_3$ (1,1) emission in the CMZ from \cite{Longmore2017}, superimposed on the CO $(J=1\to0)$ emission from \cite{Bitran+1997}. The CMZ is the concentration of molecular gas between $-1.2 \lesssim l \lesssim 2 \degree$. It is evident that the distribution of molecular gas in the CMZ is asymmetric, and there is much more emission coming from positive longitudes and velocities than from negative longitudes and velocities. Similar asymmetry can be detected in other molecular tracers such as $^{13}$CO and CS \citep[e.g.][]{Bally1988}.

\section{Methods}
\label{sec:methods}

\subsection{Numerical scheme}
\label{sec:hydro}

We have used the moving-mesh code {\sc Arepo} \citep{Springel2010}, modified to include the chemical network described in Section \ref{sec:chemistry} below. We only give a brief outline of the hydrodynamic code here and refer to the original reference for further details.

{\sc Arepo} tries to combine the strengths of Eulerian and smoothed particle hydrodynamics (SPH) codes while overcoming major weaknesses of both. It solves the fluid equations on a mesh, and in this sense is akin to Eulerian codes. The mesh is arbitrary and defined by the Voronoi tessellation of a set of discrete points. These points are not static but move with the velocity of the local flow, and in this sense the scheme is akin to SPH codes. By following the flows, the code can adjust its spatial resolution automatically and continuously, which is one of the principal advantages of SPH simulations. Moreover, there is no preferred direction unlike in most Eulerian codes. Advantages of the Eulerian methods are also retained, such as high accuracy in the treatment of shocks. {\sc Arepo} also avoids the suppression of fluid instabilities typical of SPH codes \citep{Springel2010}.

Our simulations are three-dimensional and the gas is assumed to flow in an externally imposed barred potential that rotates with a constant pattern speed $\Omega_{\rm p}$. The potential is described in Section \ref{sec:potential}. We do not include the self-gravity of the gas, nor magnetic fields. The equations of motion in an inertial frame are the conservation of mass, momentum and energy:
\begin{align}
& \frac{\pa \rho}{\pa t} + \nabla \cdot (\rho \bfv) = 0, \\
& \frac{ \pa (\rho \bfv)}{\pa t} + \nabla \cdot \left( \rho \bfv \otimes \bfv + P \bold I \right) \bfv = - \rho \nabla \Phi, \\
& \frac{\pa(\rho e)}{\pa t} + \nabla \cdot \left[(\rho e + P) \bfv \right] = \dot{Q} + \rho \frac{\pa \Phi}{\pa t},
\end{align}
where $\rho$ is the density, $\bfv$ is the fluid velocity, $P$ is the thermal pressure, $\Phi(\bfx,t)$ is the external potential, $\bold I$ is the identity matrix, $e~=~e_{\rm therm} + \Phi + {\bfv^2}/{2} $ is the energy per unit mass, $e_{\rm therm}$ is the thermal energy per unit mass and $\dot{Q}$ is a term that includes external sources of heating and cooling. We adopt the equation of state of an ideal gas,
\begin{equation}
P = (\gamma -1) \rho e_{\rm therm},
\end{equation}
where $\gamma$ is the adiabatic index. We assume for simplicity that $\gamma = 5/3$ throughout the gas; this is always accurate for atomic gas, but is also a good approximation in the H$_{2}$-dominated gas in these simulations, as very little of this gas is warm enough to excite the internal energy levels of H$_{2}$.

The above equations constitute a closed system once $\dot{Q}$ is specified. For an adiabatic gas, $\dot{Q}=0$. However, our treatment of chemistry and of heating and cooling processes affect the thermodynamics (and hence the dynamics) of the gas through the term $\dot{Q}$.

The thermal energy is related to the temperature of the gas by $e_{\rm therm} = kT / [ (\gamma -1)\mu m]$, where $\mu$ is the mean molecular weight, which is self-consistently calculated based on the chemical state of the gas, $m$ is one atomic mass unit, $T$ is the local temperature, $k$ is the Boltzmann constant. This equation is used to calculate the local temperature of the gas.

\begin{figure}
\includegraphics[width=84mm]{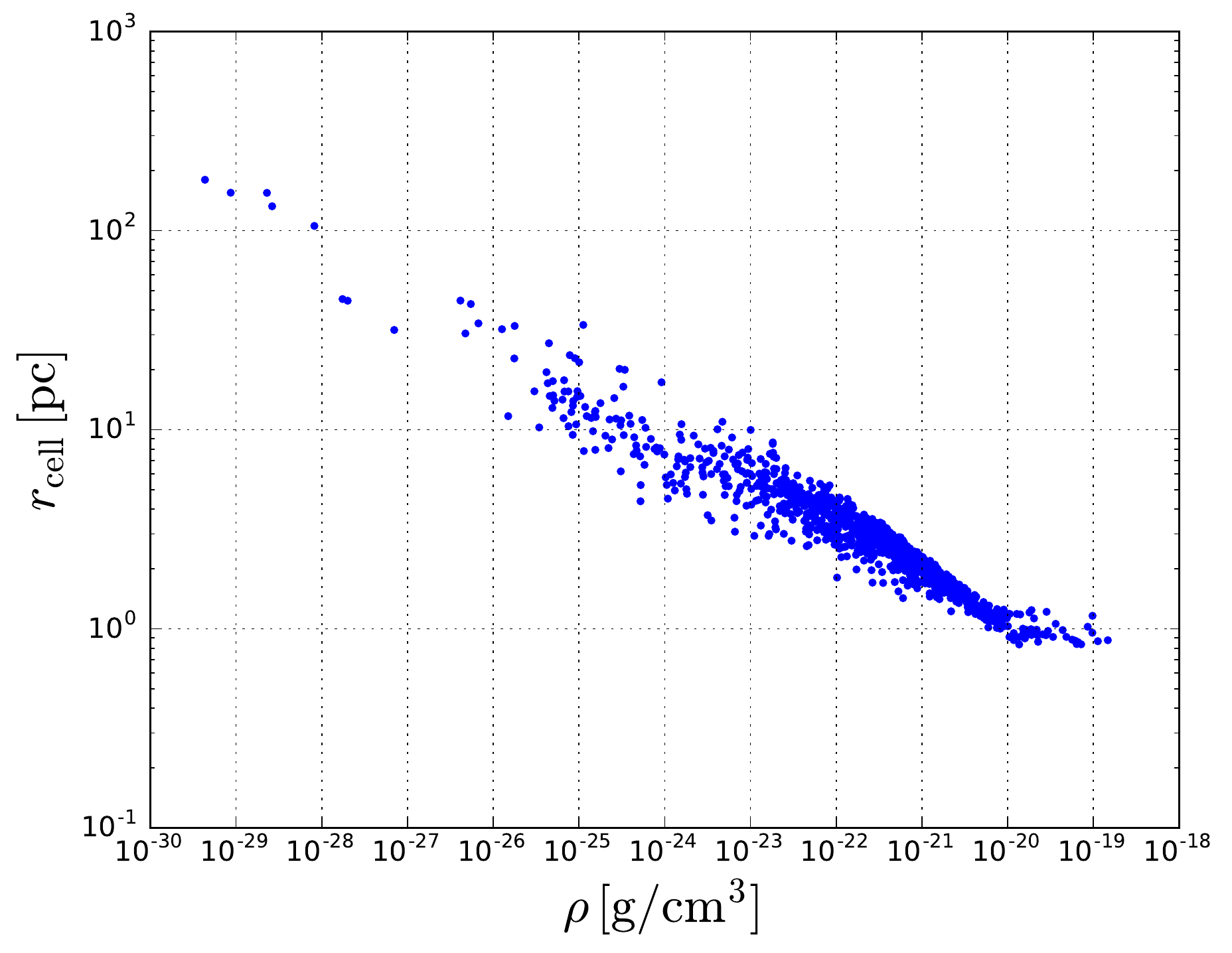}
\caption{The spatial resolution as a function of density for a random sample of cells. $r_{\rm cell}$ is defined as the length of the side of the cube that has the same volume of the cell considered.}
\label{fig:resolution}
\end{figure}

We simulate a box of $24 \times 24 \times 2 \kpc$. We use periodic boundary conditions in all directions, but the box is sufficiently wide that this choice does not affect the bar region. We use a target resolution of $100 \, \rm M_\odot$ per cell. We make use of the system of mass refinement present in {\sc Arepo}, which splits cells whose mass becomes greater than twice this target mass. Because we keep the mass of the cells approximately constant, our spatial resolution varies as a function of the local gas density. Figure \ref{fig:resolution} shows a plot of the spatial resolution as a function of density for a random sample of points. The total number of cells in each simulation is approximately $50\mhyphen60$ million.

\subsection{Initial conditions}
\label{sec:ic}

We start with gas in radial equilibrium on circular orbits in an axisymmetric potential and gradually turn on the non-axisymmetric part of the potential during the first $150\, \Myr$ to avoid transients. We initially generate $N\simeq110000$ random mesh points using a uniform 3D Poisson disc sampling \citep{Bridson2007} inside a cylindrical slab of radius $R_0 = 10\kpc$ and half-height $z_0 = 1\kpc$. We use these points to create the initial Voronoi tessellation. Thanks to the Poisson disc sampling, the volume distribution of this initial tessellation has much less scatter than if we sampled points using a standard uniform distribution. The mean radius of the initial cells is $r_0 \simeq 135 \rm \, pc$. The system of mass refinement present in {\sc Arepo} quickly creates more cells to reach the target resolution within the first couple of $\Myr$. We assign the same mass to each initial cell so that the total gas mass is $M_0\simeq10^{10}\, M_\odot$. Since the volumes of the initial cells are not exactly the same but their mass is, the initial density distribution contains moderate fluctuations over the spatial scale $r_0$. The resulting initial density distribution is therefore approximately uniform inside the cylindrical slab (both in the radial and vertical directions) with mean $\rho_0 \simeq 2.1 \times 10^{-24} \: {\rm g \, cm^{-3}}$ and standard deviation $\sigma_0 \simeq 0.5 \times 10^{-24} \: {\rm g \, cm^{-3}}$. We have tested that an initial density distribution which is exactly uniform (no random fluctuations), or that exponentially declines with radius after $R=4\, \kpc$ rather than being constant makes no difference for the results of this paper (see Appendix \ref{appendix:a}). 
 
The gas starts in atomic form, with temperature $T~=~1.5 \times 10^4 \, \rm K$ and initial fractional ionisation $x_{\rm H^{+}}~=~0.01$. We assume a hydrogen-to-helium ratio of 10:1 (by number), with all of the helium present in neutral atomic form. Carbon and oxygen are present as C$^{+}$ and O, respectively. We adopt elemental abundances of carbon and oxygen (by number, relative to hydrogen) of $x_{\rm C, tot} = 1.4 \times 10^{-4}$ and $x_{\rm O} = 3.2 \times 10^{-4}$, respectively \citep{sem00}. These initial conditions correspond to a value of $\mu=1.26$ and $\cs\equiv \sqrt{k T / \ (\mu m)}=10\kms$. For simplicity, we assume that the gas has solar metallicity, independent of Galactocentric radius.

\subsection{The potential}
\label{sec:potential}
We use the same potential as \cite{Ridley+2017}. This choice allows us to make direct comparisons between the results of the present study and this earlier work. Any changes in the gas behaviour must be due to the fact that the present simulations are three- rather than two-dimensional and are performed with a detailed treatment of the thermal evolution of the gas rather than the assumption of isothermality. Here, we briefly recap the characteristics of the potential employed and refer to \cite{Ridley+2017}  for further details.

We use a realistic Milky Way potential that is the sum of four components: bar, bulge, disc, and halo. The axisymmetric part is derived from the work of \cite{McMillan2017}, whose potential is created to fit observational constraints and to be consistent with expectations from theoretical modelling of the Milky Way as a whole. The bar and the bulge are built to be consistent with observational constraints from near-infrared photometry \citep{Launhardt+2002} and with dynamical constraints on the quadrupole of the bar \citep{SBM2015c}.

We expand the potential in the plane of the Galaxy in multipoles, so
\begin{equation}
\Phi(R,\phi) = \Phi_0(R) + \sum_{m=1}^{\infty} \Phi_m(R) \cos\left(m \phi + \phi_m\right),
\end{equation}
where $\phi_m$ are constants and $\{R,\phi,z\}$ denote cylindrical polar coordinates.

\begin{figure}
\includegraphics[width=84mm]{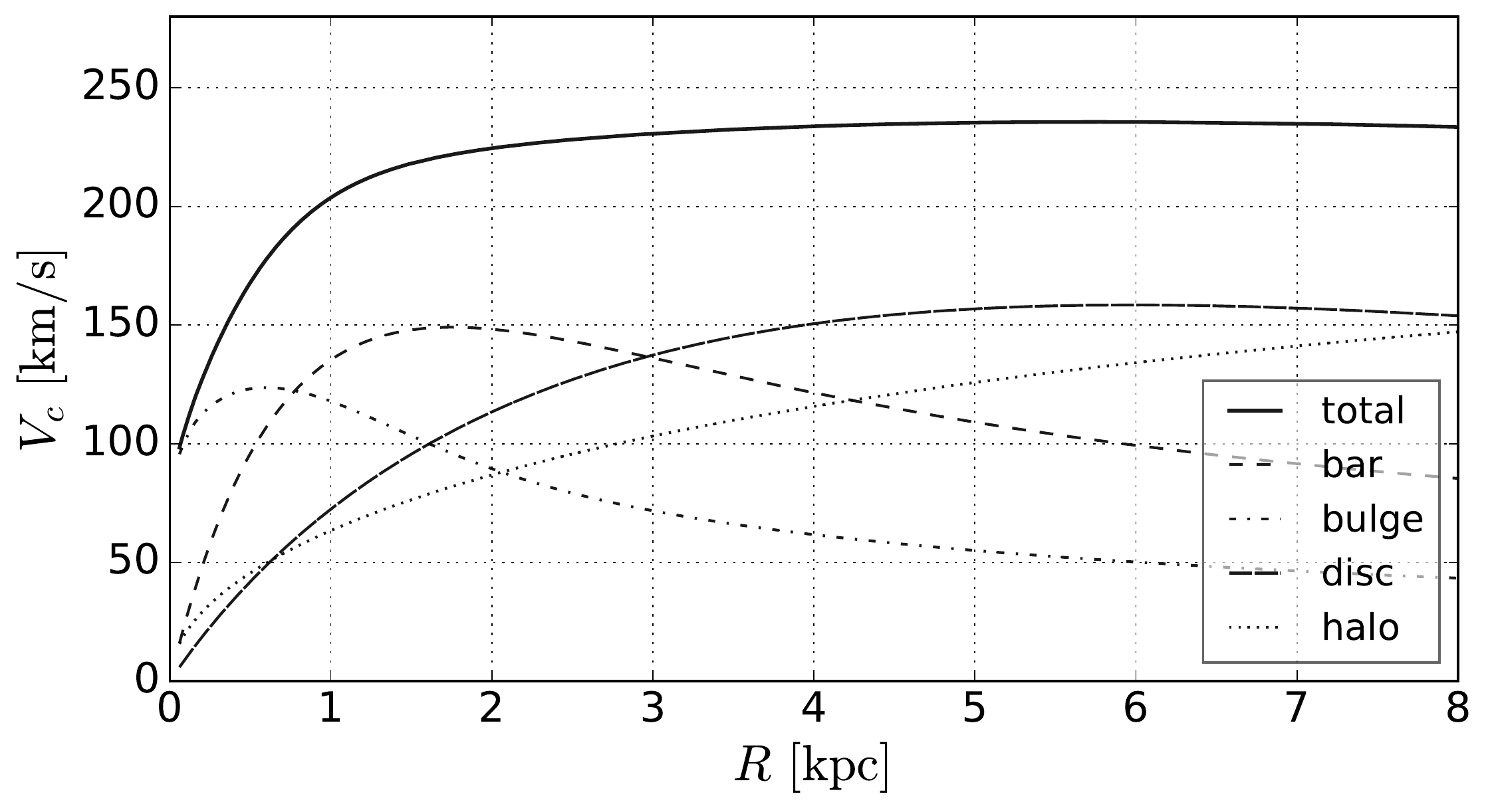}
\caption{The circular velocity curve for the potential used in this paper. The separate contributions from bar, bulge, disc and halo are also shown.}
\label{fig:vc}
\end{figure}

Fig. \ref{fig:vc} shows the ``circular speed curve'' of the potential and the contributions from each component separately. Note that our definition is based on the axisymmetric part of the potential:
\begin{equation}
V_c(R) \equiv \sqrt{ R \frac{\di \Phi_0}{\di R} }.
\end{equation}
Since the gas undergoes strong non-circular motions in the region dominated by the bar, which for our Galaxy corresponds approximately to the region within Galactocentric radius $R = 3 \, \kpc$ \citep[e.g.][]{BM}, the ``circular speed'' can be significantly different from the speed of the gas obtained in simulations, or observed in the Galaxy, at the same radii \cite[e.g.][]{Binney++1991}.

\begin{figure}
\includegraphics[width=84mm]{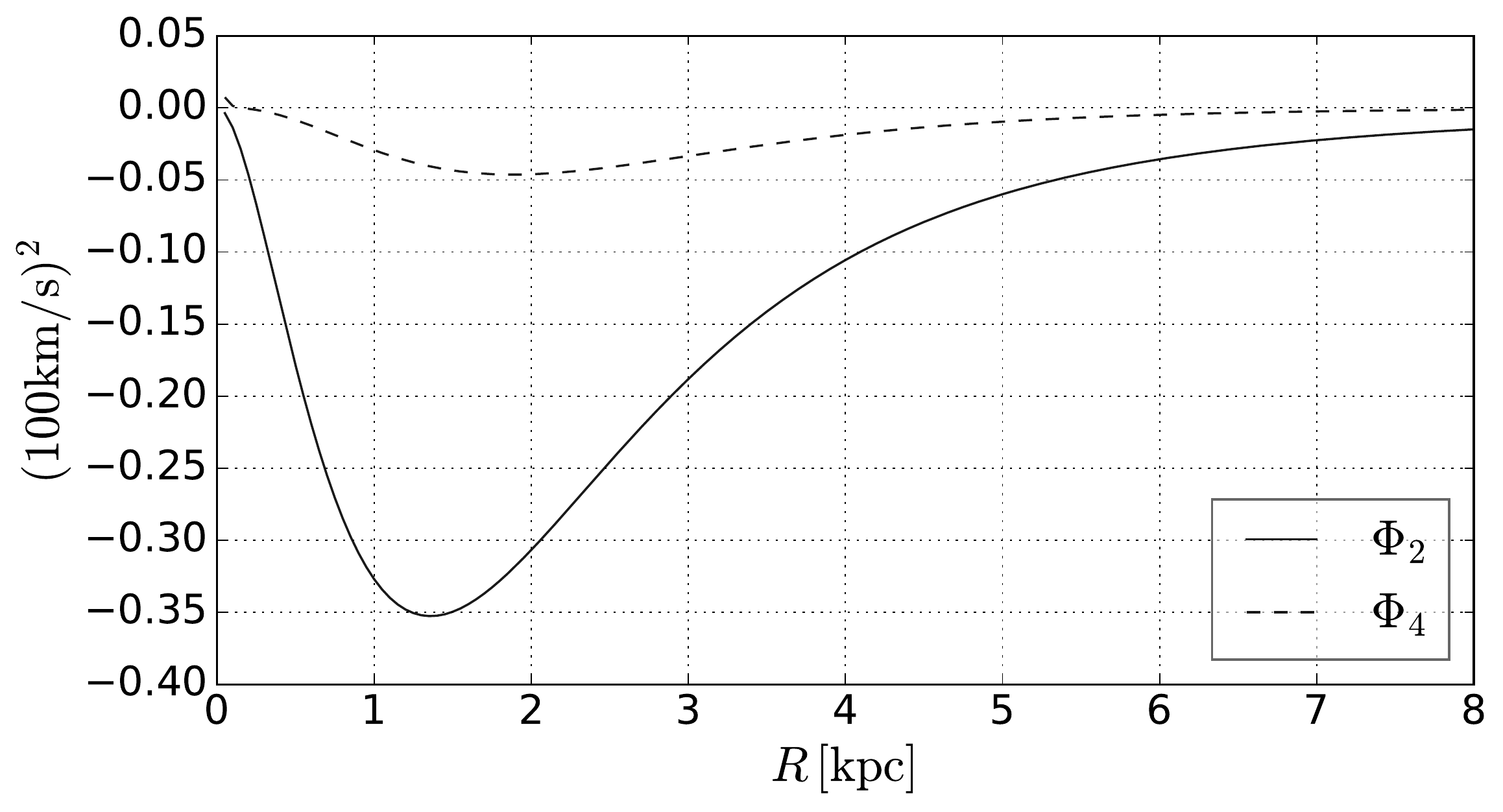}
\caption{The quadrupole $\Phi_2$ and octupole $\Phi_4$ of the potential used in this paper.}
\label{fig:multipoles}
\end{figure}

Fig. \ref{fig:multipoles} shows the quadrupole, $m=2$, and the octupole, $m=4$, of the potential used in this paper. These are generated by the bar, which is the only non-axisymmetric component in our potential. The bar is assumed to rotate rigidly with a constant pattern speed of $\Omega_{\rm p} = 40\kms$, consistent with recent estimates \citep[e.g.][]{SBM2015c,WeggGerhard2015,Perez+2017}. This places the Inner Lindblad Resonance at $R_{\rm ILR}=1.23\, \kpc$, corotation at $R_{\rm CR}=5.89\, \kpc$ and the Outer Lindblad Resonance at $R_{\rm OLR}=9.75\, \kpc$. Note however that the value of the pattern speed is uncertain. It was until recently debated in the literature whether it should be ``low'',  $\Omega_{\rm p} \lesssim 40\kms$, or ``high'', $\Omega_{\rm p} \simeq 60\kms$, with the community now generally preferring a ``low" value \citep[see for example the review of][]{BlandHawthornGerhard2016}.

\subsection{Chemistry, heating \& cooling}
\label{sec:chemistry}
We follow the chemical evolution of the gas using a simplified network for hydrogen, carbon and oxygen chemistry based on the work of \citet{gm07a,gm07b} and \citet{nl97}. In this scheme, changes in the chemical composition and heating \& cooling of the gas are deeply coupled. The same network has previously been used for a number of other applications, such as modelling the small-scale evolution of molecular clouds in a CMZ-like environment \citep{Bertram+2015}, or studying the chemical and dynamical evolution of gas passing through spiral arms in a large-scale Galactic spiral potential \citep{Smith+2014}. We summarise here the main points and refer the reader to the cited papers for further details.

Our scheme proceeds along the following lines. At any given time, the state of each {\sc Arepo} mesh cell is completely specified by the following basic quantities:
\begin{itemize}
\item The total gas density $\rho$.
\item The fractional abundances of each chemical component $\{x_i\}$, where $i~=~\{ \rm H, H_2, H^+, He, C^{+}, CO, O, e^{-} \}$. These can be used to derive the corresponding number densities $\{n_i\}$. The total density satisfies $\rho~=~\sum_i \rho_i$, where $\rho_i$ are the mass densities of the individual components.
\item The velocity $\bfv$ of the cell.
\item The thermal energy per unit mass, $e_{\rm therm}$, from which we can derive the temperature $T$ of the gas.
\end{itemize}
To solve the governing equations for the chemical abundances, we use operator splitting to treat separately changes due to the advection of the partial densities with the local velocity $\bfv$ and changes due to reactions creating or destroying chemical species. Similarly, when solving the energy equation, we treat the evolution due to radiative and chemical heating and cooling separately from the evolution due to adiabatic compression and rarefaction or shocks. Our treatment of advection, $p{\rm d}V$ heating/cooling etc. is exactly the same as in the standard version of {\sc Arepo} \citep{Springel2010} and so we do not discuss it further here.

To treat the reaction terms for the chemical species and the radiative and chemical heating and cooling terms for the internal energy, we proceed as follows. We begin by calculating the  
following derivative quantities for each cell:
\begin{itemize}
\item The dust grain number density $n_{\rm dust}$ and temperature $T_{\rm dust}$. The former is calculated assuming a constant dust-to-gas ratio, and ignoring any change in the dust properties (composition, size distribution etc.) with position in the Galaxy. To calculate the dust temperature, we follow the procedure described in Appendix A of \citet{gc12}. We assume that the dust is in thermal equilibrium, and that its temperature is set by the balance between heating due to absorption of photons from the interstellar radiation field (ISRF) and energy transfer in collisions with gas particles, and cooling due to the thermal radiation of the dust grains. 
\item The intensity of the ISRF. This is obtained starting from the value prescribed locally by a given profile $G(R)$, defined relative to the standard value $G_0$ from \cite{draine78}, and then attenuating it to account for the effects of dust extinction and H$_{2}$ self-shielding. This is done by first using the {\sc Treecol} algorithm \citep{clark12} to compute a $4\pi$ steradian map of the H$_{2}$ and dust column densities seen by the {\sc Arepo} mesh cell, and then using these values to compute the angle-averaged shielding factors due to dust and H$_{2}$. As in \citet{Smith+2014}, we only account for gas and dust within 30~pc of each mesh cell when computing the shielding, which allows us to properly capture the influence of the local gas distribution but ensures that we do not overestimate the shielding present along lines of sight passing through the mid-plane of the disc. We experiment with different choices of the function $G(R)$ as detailed in Sect. \ref{sec:summarysims}.
\item The intensity of the cosmic-ray ionisation rates. For atomic hydrogen, we simply scale the rate $\zeta_{\rm H}$ proportionally to the same profile $G(R)$ used for the ISRF, without any attenuation since cosmic rays penetrate even into dense molecular clouds. The scaling factor is fixed such that the rate corresponding to $G_0$ is $3 \times 10^{-17} \: {\rm s^{-1}}$.
For the other cosmic ray processes, we then scale their rates relative to the hydrogen rate using values taken from \cite{wood07}. 
\end{itemize}

The second task performed by the code once the above quantities are known is updating the chemical composition of the ISM from one time-step to the next. The slowest chemical reaction in our network is H$_2$ formation, which has a time-scale of roughly $1 \rm{Gyr}/n$, where $n$ is the total number density in units of cm$^{-3}$ \citep{hm89}. In dense gas, this is usually much shorter than the hydro timescale, but in more diffuse gas it can be of the order of several Myr. Hence the ``chemical'' time-step used to keep track of the chemical networks is usually shorter than the time-step used in the hydrodynamic part of the code. The reactions included in the network are listed in Table~\ref{tab:chem}. For the most part, these are self-explanatory, but our treatment of CO formation deserves further comment. Following \citet{nl97}, we assume that C$^{+}$ is converted to CO without directly tracking all of the intermediate reactions (and intermediate species) that are actually involved (c.f.\ the more complicated model in \citealt{glo10}, which does attempt to track them). The rate limiting step in the formation of CO is generally the formation of the initial radical or molecular ion -- in this case, CH$_{2}^{+}$ -- and we assume that only two outcomes are possible after this: either the CH$_{2}^{+}$ ion initiates a chain of reactions that ends in the formation of a CO molecule, or it is instead photodissociated by the ISRF. \citet{gc12} compared this simple prescription with several more complicated approaches and showed that it does a reasonable job of predicting which gas will be CO-rich and which will be CO-poor, although it tends to overproduce CO at moderate visual extinctions. Its accuracy is sufficient for our current purposes given that with our resolution ($\sim 100 M_\odot$/cell) details of the internal structure of individual molecular clouds remain unresolved.

At the same time that the code is evolving the chemistry, it is also updating the thermal energy by computing the various heating \& cooling processes that contribute to $\dot{Q}$ and using these to determine $e_{\rm therm}$. The heating and cooling processes are modelled using the detailed atomic and molecular cooling function introduced in \citet{glo10} and updated in \citet{gc12}. Table~\ref{tab:cool} gives a full list of the processes that are included in this cooling function. Note that in this scheme the cooling function does not only depend on $\rho$ and $T$, but also on the abundances $\{x_i\}$, on the value of the attenuated ISRF, on the value of the cosmic ray ionisation rates and on the temperature of gas and dust. These yield instantaneous rates for all the chemical reactions included and hence the related instantaneous amounts of heating \& cooling. This is done with a sufficiently fine ``chemical'' time-step that all of the important chemical reactions are well resolved. In practice, we can distinguish between a few regimes. In gas with a low dust extinction, the heating is dominated by the photoelectric effect from the assumed external ISRF, and the cooling is dominated in warm gas by the collisional excitation of the permitted transitions of atomic hydrogen (``Lyman-$\alpha$ cooling''), atomic helium and heavier elements, and in cooler gas ($T < 8000$~K) by fine structure emission from C$^{+}$ and O. In well-shielded regions, on the other hand, cosmic rays dominate the heating, and CO and dust dominate the cooling.

\begin{figure}
\includegraphics[width=0.48\textwidth]{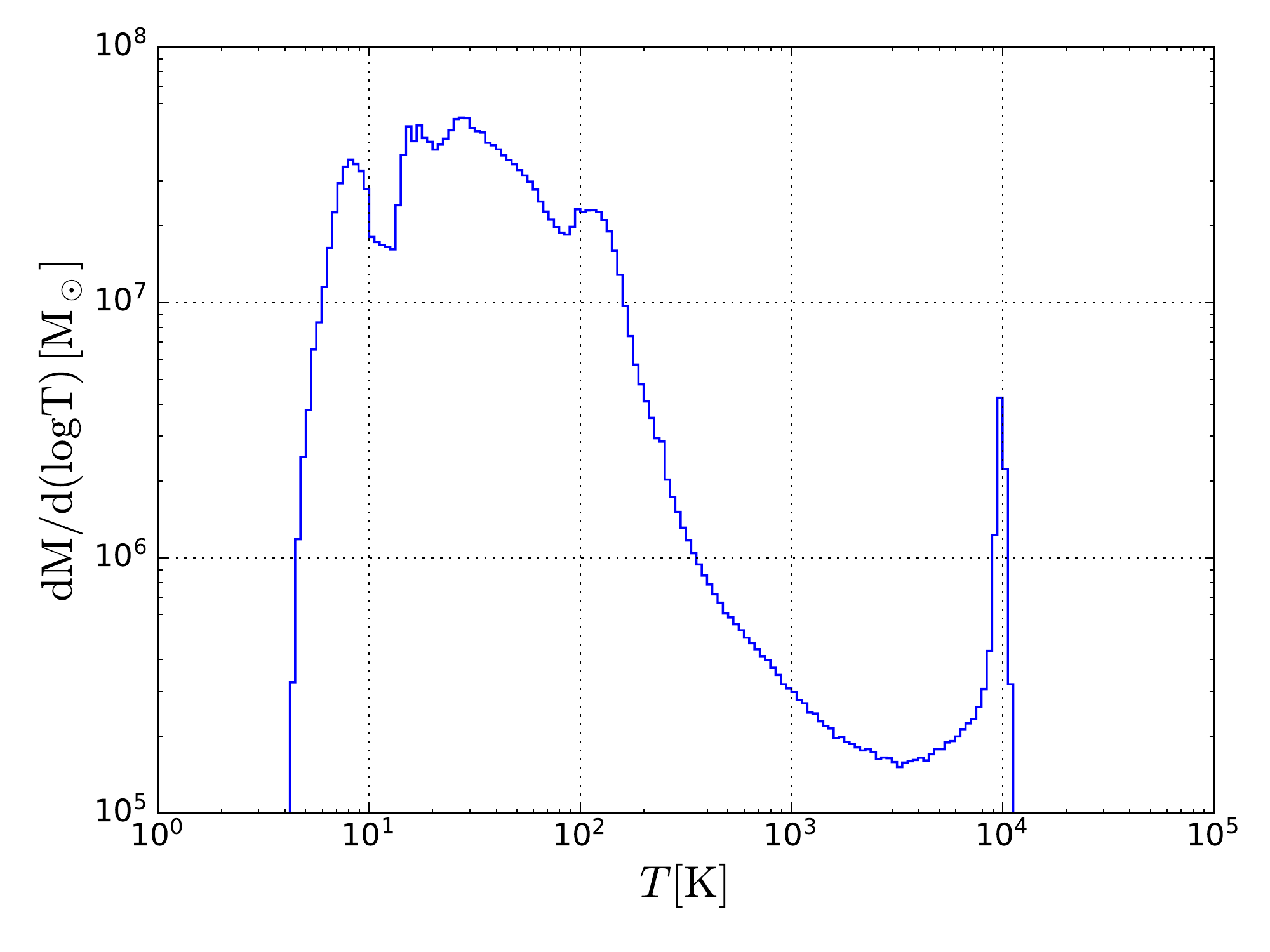} 
\caption{Histogram showing the total amount of mass contained in each temperature bin. Binned are all grid cells of the snapshot at $t = 181.1 \, \Myr$ of the variable simulation (c.f.\ Fig. \ref{fig:rho2}) with $R \leq 3\kpc$. The separation of the gas into distinct cold and warm thermal phases is clear. Note that bin sizes are logarithmic.}
\label{fig:T}
\end{figure}

Finally, we note that owing to the weak temperature dependence of the dominant heating and cooling processes over a broad range of gas temperatures, the cooling function used in this work gives rise to a medium which is susceptible to the thermal instability \citep{Field1965}. We thus generically find that the gas in our simulations tends to settle into a two-phase medium of the type envisaged by \cite{Field+1969}. In this model, two stable phases can co-exist in pressure equilibrium, one cold and dense ($T\sim100 \, \rm K$), and one warm and diffuse ($T\sim10^4 \, \rm K)$, which we can identify with the cold neutral medium (CNM) and the warm neutral medium (WNM), respectively. Fig. \ref{fig:T} illustrates this dichotomy by showing a histogram of the mass vs. temperature for a snapshot of our simulations (see also the analysis in \citealt{GloverSmith2016}, who examine the temperature distribution of the gas in a set of high resolution galactic simulations performed using the same treatment of chemistry and cooling function as in this work). The existence of two distinct thermal phases leads to significant differences in the dynamics of the gas in our simulations in comparison to the isothermal simulations presented in \cite{Ridley+2017}, as we explore in more detail below.

\begin{table}
\caption{Reactions in the chemical model \label{tab:chem}}
\begin{tabular}{lc}
Reaction & Reference(s) \\
\hline
${\rm H + e^{-}} \rightarrow {\rm H^{+} + e^{-} + e^{-}}$ & \citet{abel97} \\
${\rm H^{+} + e^{-}} \rightarrow {\rm H + \gamma}$ & \citet{fer92} \\
${\rm H^{+} + e^{-} + grain} \rightarrow {\rm H + grain}$ & \citet{wd01} \\
${\rm H + c.r.} \rightarrow {\rm H^{+} + e^{-}}$ & See text \\
${\rm H + H + grain} \rightarrow {\rm H_{2} + grain}$ & \citet{hm89} \\
${\rm H_{2} + H} \rightarrow {\rm H + H + H}$ & \citet{ms86} -- low $n$; \\
& \citet{msm96} -- high $n$ \\
${\rm H_{2} + H_{2}} \rightarrow {\rm H + H + H_{2}}$ & \citet{mkm98} -- low $n$; \\
& \citet{sk87} -- high $n$ \\
${\rm H_{2} + e^{-}} \rightarrow {\rm H + H + e^{-}}$ & \citet{tt02} \\
${\rm H_{2} + \gamma} \rightarrow {\rm H + H}$ & \citet{db96}$^{*}$ \\
${\rm H_{2} + c.r.} \rightarrow {\rm H + H^{+} + e^{-}}$ & See text \\
${\rm H_{2} + c.r.} \rightarrow {\rm H + H}$ & See text \\
${\rm C}^{+} + {\rm H_{2}} \rightarrow {\rm CH}_{2}^{+} + \gamma$ & \citet{nl97} \\
${\rm CH_{2}^{+} + O} \rightarrow {\rm CO + products}$ & \citet{nl97} \\
${\rm CH}_{2}^{+} + \gamma \rightarrow {\rm C}^{+} + {\rm H_{2}}$ & \citet{nl97}$^{*}$ \\
${\rm CO} + \gamma \rightarrow {\rm C + O}$ & \citet{vis09}$^{*}$ \\
\hline
\end{tabular}
\\ $^{*}$ The rates of these processes are scaled with the strength of the local ISRF and
the effective dust extinction as described in the text.
\end{table}

\begin{table}
\caption{Processes included in the cooling function \label{tab:cool}}
\begin{tabular}{lc}
Process & Reference(s) \\
\hline
H collisional excitation & \citet{black81,cen92} \\
H collisional ionisation & \citet{abel97} \\
H$^{+}$ recombination cooling & \citet{fer92}  -- gas phase; \\
& \citet{wol03} -- grains \\
H$_{2}$ ro-vibrational lines & \citet{ga08} \\
H$_{2}$ collisional dissociation & See Table~\ref{tab:chem} \\
H$_{2}$ formation heating & \citet{hm79} \\
H$_{2}$ photodissociation heating & \citet{bd77} \\
H$_{2}$ pumping by UV & \citet{bht90} \\
Atomic resonances lines$^{*}$ & \citet{sd93} \\
C$^{+}$, O, Si$^{+}$ fine structure lines & \citet{glo10} \\
CO rotational lines & \citet{nk93}; \\
& \citet{nlm95} \\
Gas-grain energy transfer & \citet{hm89} \\
Photoelectric heating & \citet{bt94}; \\
& \citet{wol03} \\
Cosmic ray heating & \citet{gl78} \\
\hline
\end{tabular}
\\ $^{*}$ Emission from He and heavier elements at $T > 10^{4} \: {\rm K}$.
\end{table}

\subsection{Summary of the simulations run} \label{sec:summarysims}

\begin{figure}
\includegraphics[width=84mm]{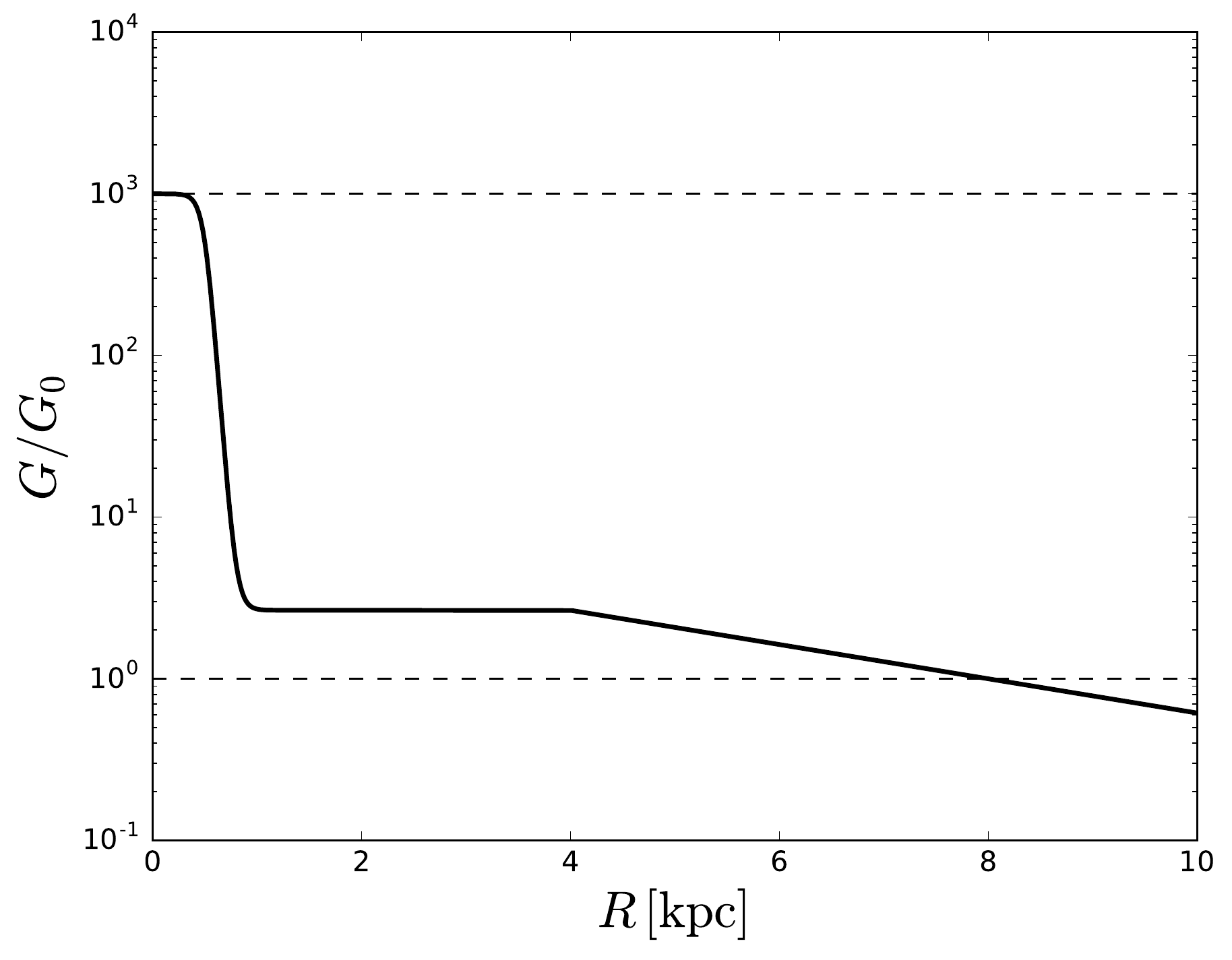}
\caption{Profiles of the adopted ISRF and cosmic ray ionisation rate for the three simulations considered in this paper. The bottom and top dashed lines correspond to the ``low'' and ``high'' simulations respectively, while the full line corresponds to the ``variable'' simulation. $G_{0}$ corresponds to the ISRF strength given in \citet{draine78}. We assume that the cosmic ray ionisation rate scales linearly with $G$ and adopt a value $\zeta_{\rm H} = 3 \times 10^{-17} \: {\rm s^{-1}}$ when $G / G_{0} = 1$.}
\label{fig:ISRF}
\end{figure}

We have run three simulations in total, which differ only in the assumed radial profile, $G(R)$, of the ISRF and cosmic ray ionisation rate. Since the value of $G(R)$ is quite uncertain far from the Solar neighbourhood \citep[e.g.][]{Wolfire+2003}, it is necessary to test the robustness of our results agains this uncertainty. Figure \ref{fig:ISRF} shows the profiles corresponding to the three simulations,
which we named ``low'', ``variable'' and ``high''. In the ``low'' and ``high'' simulations, which correspond to the bottom and top dashed lines, the ISRF is assumed to be constant. The ``low'' line corresponds to the value measured in the Solar neighbourhood, $G_0$ \citep[e.g.][]{draine78}. The ``high'' line corresponds to a plausible estimate of the field present in the central 100~pc
region in the CMZ, $1000G_0$ \citep[e.g.][]{Clark+2013}. Hence these correspond to limiting values for the average ISRF in the bar region. The ``variable'' profile is constructed as one possible way of interpolating between the high value inferred for the inner CMZ and the profile at $R > 3 \, {\rm kpc}$ given in \citet{Wolfire+2003} and is given by
\begin{equation}
\frac{G(R)}{G_0} = \begin{cases} \frac{1000 - {\rm e}}{1 + \exp{\left( \frac{R - 0.5\kpc }{0.05\kpc}\right)}} + {\rm e} \quad &{\rm if} \quad R\leq4\kpc \\ \exp{\left(-\frac{R - R_0}{4 \kpc}\right)} \quad &{\rm if} \quad R>4 \kpc \end{cases},
\end{equation}
where $R_0 = 8\kpc$ is the Sun-Galactic centre distance. Note that we do not linearly interpolate in the regime $0.5 < R < 3 \: {\rm kpc}$ as there is little ongoing star formation in the Galaxy at these radii \citep[e.g.][]{MorrisSerabyn1996},  so a simple linear interpolation with radius may dramatically over-estimate the ISRF strength at these radii.

\subsection{Producing synthetic $(l,b,v)$ data cubes}
\label{synth}

We adopt a very simple projection procedure to compute the predicted $(l,b,v)$ distributions from each simulation snapshot. In all of our projections, we assume that the Sun is undergoing circular motion at a radius $R_0=8\kpc$, at $z=0$ with speed $v_\odot = 220\kms$ and that the angle between the bar major axis and the Sun-Galactic Centre line is $20\degree$, consistent with current estimates \citep[e.g.][]{BlandHawthornGerhard2016}.

We bin each {\sc Arepo} cell as a point in an $(l,b,v)$ data cube with weights proportional to the mass of the component under consideration (H{\sc I} or CO) and inversely proportional to the square of its distance from the Sun. The resolution of this binning is $\Delta l=0.025\degree$, $\Delta b = 0.025\degree$ and $\Delta v=0.25\kms$. 

This approach assumes that the contribution made by each cell to the observed brightness temperature is directly proportional to the mass of the tracer in the cell. This assumption is valid for H{\sc I}, independent of the density or temperature of the gas, provided that the H{\sc I} is optically thin. Since we do not expect much of the H{\sc I} in the Galaxy to be optically thick, the resulting emission map will be very similar to the one that we would recover from a full radiative transfer calculation \citep[e.g.][]{KalberlaKerp2009}. In the case of CO, on the other hand, we expect much of the emission to be optically thick and therefore this approximation will generally overestimate the emission. However, for our current purposes, we are not particularly interested in the intensity of CO emission coming from a particular $(l,b,v)$, but rather in whether there is {\em any} emission coming from this direction at this velocity. Our approach does do a good job of predicting which points in $(l,b)$ or $(l,v)$ have associated emission, and is therefore useful for characterising the overall morphology and large-scale features seen in the molecular gas distribution, even in cases where it does not accurately predict the intensity of the emission.

\section{Gas dynamics}
\label{sec:results}

\subsection{Dynamics in the $(x,y)$ plane} \label{sec:xy}

Figure \ref{fig:rho1} shows the temporal evolution of the total gas surface density for the simulation with a spatially variable ISRF. The gas is initially on circular orbits and the bar is turned on gradually during the first $150 \Myr$. Figure \ref{fig:rho2} shows the gas surface density of different components for the snapshot at $t=181.1\Myr$. From top to bottom, the panels show the total, CO, H$_2$ and H surface densities, respectively. Fig \ref{fig:vfield} shows the velocity field in the frame corotating with the bar for the same snapshot. In all these images, material rotates in the clockwise direction and the (stellar) bar major axis is horizontal.

\subsubsection{$x_1$ and $x_2$ orbits}

Closed orbits of the $x_1$ and $x_2$ families form the basis to understand the gas flow in a barred potential in much the same way circular orbits form the basis to understand the gas flow in an axisymmetric disc galaxy \citep[e.g.][]{Binney++1991}. Most prograde ballistic orbits in a bar potential, open or closed, can be interpreted as epicyclic excursions around these two families of stable closed orbits. $x_1$ orbits are strongly elongated in the direction of the bar, while $x_2$ orbits are mildly elongated perpendicular to the bar. The top panel in Fig. \ref{fig:rho2} shows examples of these two types of orbits for the potential employed in this paper.

Schematically, the gas flow can be understood as follows. Material in the outer region follows $x_1$ orbits, while dissipative processes cause it to slowly drift inwards along a sequence of such orbits. If the potential possesses an Inner Lindblad Resonance, as in our case, $x_1$ orbits become self-intersecting below some critical energy. Therefore the gas cannot continue to follow $x_1$ orbits and it transits within a dynamical time from the $x_1$ to the $x_2$ orbits. The transition happens through the formation of two almost straight ``bar shocks'' (see Fig. \ref{fig:rho2}), along which the gas plunges almost radially (c.f. the velocity field in Fig. \ref{fig:vfield}) towards the centre, where it settles onto a disc/ring like structure of material that follows approximately $x_2$ orbits. In the interpretation of \cite{Binney++1991} this $x_2$ disc/ring structure corresponds to the CMZ (see left panel in the second row and all right panels in Fig. \ref{fig:rho2}). Hydrodynamical models refine this picture by properly taking into account the effects of gas pressure neglected in the ballistic approximation (see for example \citealt{SBM2015a,SBM2015b} for a detailed comparison of the results of a simple two-dimensional hydro simulation vs closed orbits).

Despite the addition of the chemical network and the third dimension, the large-scale dynamics in and around the bar remains similar to that of the isothermal model of \cite{Ridley+2017}, and the schematic description above remains valid at zeroth-order. The main difference is that, in contrast to the isothermal simulations in which the CMZ disc has a smooth density distribution, here it is broken up into several large molecular clouds (see Fig. \ref{fig:rho2}). The centres of mass of these large clouds closely follow closed ballistic $x_2$ orbits, but on top of this basic motion there are significant excursions and spinning motions (see also Section \ref{sec:CMZ} and in particular Fig. \ref{fig:lv3}).

\subsubsection{Wiggle instability, thermal instability and episodic CMZ bombardment} \label{sec:bomb}

There are three distinct dynamical mechanisms that play a major role in shaping the simulated CMZ morphology.

\subparagraph{Wiggle instability.} The two bar shocks that form in a gas flow in barred potentials are unstable. They are subject to a corrugation instability of shock fronts also known as ``wiggle instability'' \citep{WadaKoda2004,KimKimKim2014,SSSTK}. This instability causes the shocks to periodically break apart and reform, resulting in an unsteady and asymmetric flow of gas down the shocks onto the CMZ. The physical mechanism behind the instability is periodicity: every time some inhomogeneities cross a shock in the course of Galactic rotation, they are significantly amplified, quickly leading to instability (\citealt{SSSTK}; see also \citealt{DobbsBonnell2006}). For this reason, the instability may be more appropriately called the ``Periodic Shock Instability". The wiggle instability was present in the isothermal simulations of \cite{SBM2015a} and \cite{Ridley+2017}, which is what led them to suggest it as a key ingredient in the explanation of the CMZ asymmetry. 

\subparagraph{Thermal instability.} The thermal instability of \cite{Field1965} appears to make the wiggle instability much more powerful for two reasons. First, it creates density inhomogeneities. Our density distribution at $t=0$ is smooth and uniform (except for some small random noise due to the way the initial conditions are generated, see Section \ref{sec:ic}) and the gas is all in the warm phase at $T \simeq 1.5 \times 10^4 \, \rm K$. Scattered clumps of dense cold material are stochastically formed throughout the Galactic plane during the first few megayears due to the thermal instability (e.g. see the CO distribution in Fig. \ref{fig:rho2}). These are embedded into a warmer, more diffuse phase much like in the two-phase medium envisaged by \cite{Field+1969}. These clumps are then stirred and sheared by Galactic rotation, which causes them to aggregate into bigger large-scale patterns that resemble the bar-driven density waves studied in \cite{SBM2015b}. These density waves can also be subject to the wiggle instability if their amplitude is strong enough for them to essentially become spiral shocks \citep[c.f. figure 8 of][]{SBM2015c}. Since the wiggle instability works by amplifying inhomogeneities, this makes its development faster and more violent.

Second, the cold phase created by the thermal instability has a very low sound speed, of the order of $\lesssim 1\kms$. The warm phase in contrast has a sound speed closer to $\sim10 \kms$, a typical value assumed in isothermal simulations.\footnote{Note however that the sound speed of an isothermal simulations is not meant to be related to the microscopic temperature of gas in a disk galaxy, but is instead a phenomenological sound speed that takes into account in a simple way the turbulent pressure of the interstellar medium  \citep[e.g.][]{Roberts1969,Cowie1980}. The ``temperature'' of the isothermal assumption is therefore related to the velocity dispersion of clouds rather than a microscopic temperature.} The wiggle instability is known to have a very strong dependence on sound speed, and colder gas is more prone to the instability since it induces stronger shocks \citep{KimKimKim2014}. Hence, the presence of a cold phase makes the wiggle instability stronger than in isothermal simulations, which effectively behave more like the warm phase.

We have just described how the thermal instability enhances the wiggle instability. It is also worth mentioning that the interplay between the two instabilities is not unidirectional: local density enhancements created by the wiggle instability are the formation points of molecular clouds due to the thermal instability. This delineates a complicated and self-sustaining interplay between the two mechanisms eventually leading to the clumpy nature of the ISM observed in the simulation.

\subparagraph{CMZ bombardment} The two instabilities combined have the result that the CMZ is fed intermittently with shocked gas clouds that fall in rapidly from the tip of the bar and crash into the $x_2$ disc/ring. This bombardment significantly affects the dynamics of the CMZ. A large clump of material falling onto the central disc can disrupt it, causing it to break into several discrete molecular clouds. For example, it can be seen from the left panel in the second row of Fig. \ref{fig:rho2} that the CO distribution in the CMZ is made up of several distinct molecular cloud complexes with a  ``pointy'' and ``rippled'' appearance. The centre of mass of each cloud follows well a closed $x_2$ orbit, but there are significant excursions on top of this basic motion and the clouds are continuously stirred and sheared, which may be linked to the turbulent structure observed in CMZ clouds \citep[e.g.][]{Federrath+2016}. They also have a significant degree of spin. Figure \ref{fig:lv3} below follows one of these clouds for about half an orbit. 

\begin{figure*}
\includegraphics[width=1.0\textwidth]{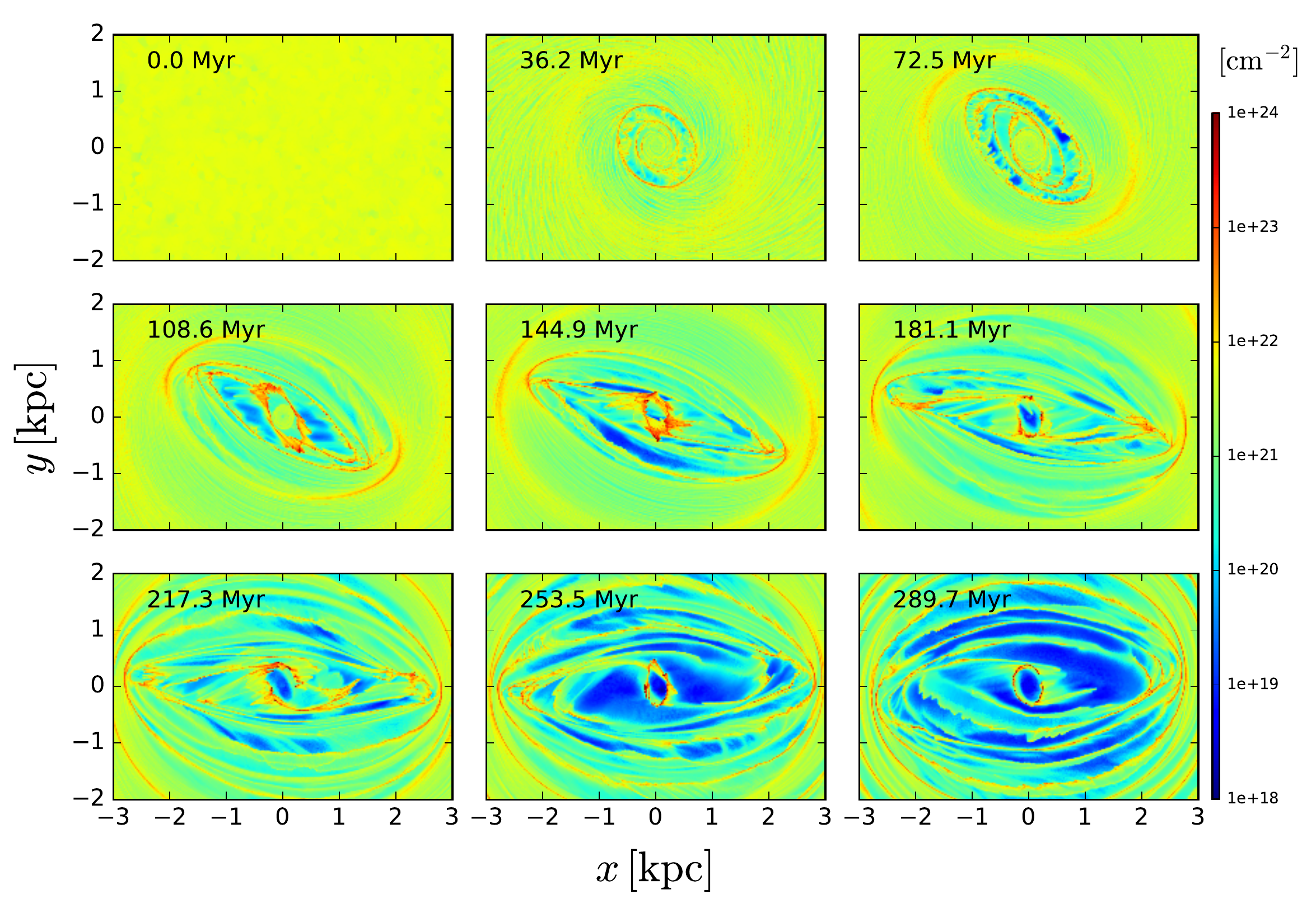}
\caption{Temporal evolution of the total particle number surface density for the simulation with a variable ISRF. The gas rotates clockwise and the major axis of the stellar bar is horizontal in all panels. The gas is initially on circular orbits and the bar is turned on gradually during the first $150 \Myr$. When the bar is weak the gas pattern is not aligned with the underlying potential.
}
\label{fig:rho1}
\end{figure*}

\begin{figure*}
\includegraphics[width=1.0\textwidth]{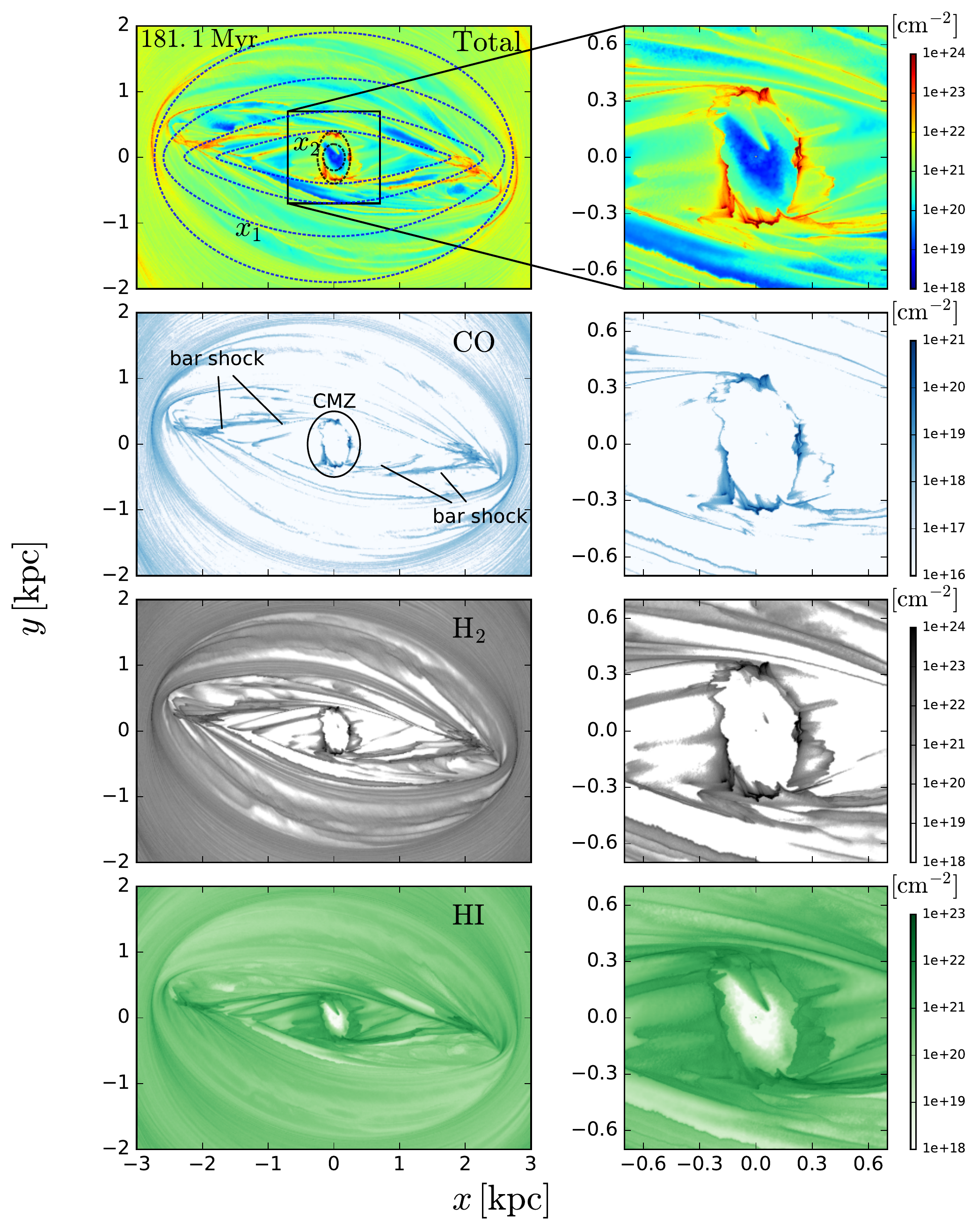}
\caption{Gas surface density at $t=181.1\, \Myr$ for the simulation with a variable ISRF. From top to bottom the panels show the following components: total density, CO, H$_2$, and H. Superimposed in the top panel are traces of $x_1$ (outermost four) and $x_2$ (innermost two) orbits.
}
\label{fig:rho2}
\end{figure*}

\begin{figure*}
\includegraphics[width=1.0\textwidth]{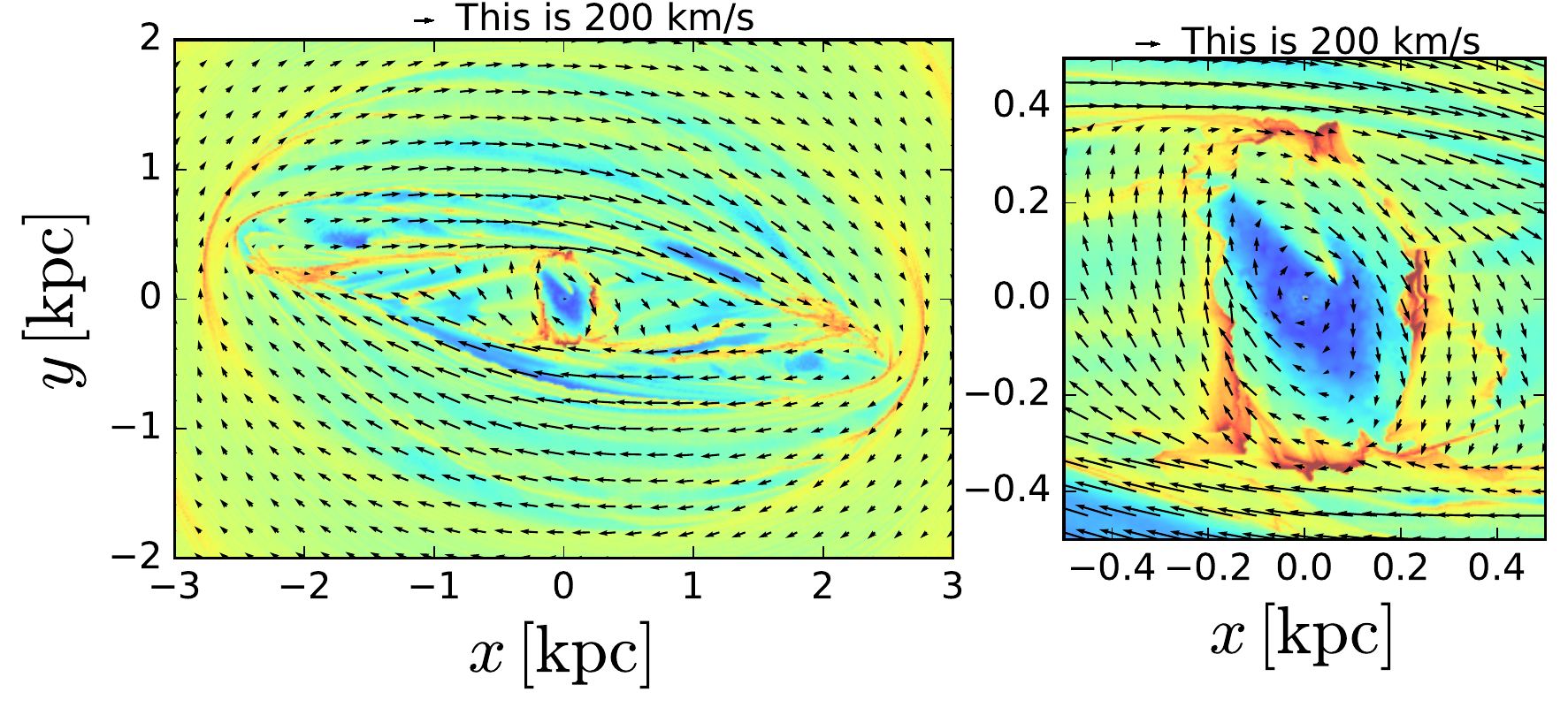}
\caption{Velocity field in the frame corotating with the bar for the same snapshot as in Fig. \ref{fig:rho2}.}
\label{fig:vfield}
\end{figure*}

Material falling from the tips of the bar does not always collide with the CMZ. Depending on the launching conditions, which are not always the same due to the unsteady flow, material either crashes into the CMZ, or overshoots, flying by it and reaching the tip of the bar on the opposite side. In our simulations we have no sinks or sources of gas taking into account processes that can replenish the gas supply, such as star formation or accretion from the Galactic halo onto the disc. Thus, we eventually run out of fresh molecular gas to feed the bar shocks and the bombardment of the CMZ decreases. Indeed, in the snapshots at $t=253.5\, \Myr$ and $t=289.7\, \Myr$ in Fig. \ref{fig:rho1} it can be seen that the innermost non self-intersecting $x_1$ orbits are bluer than in other panels and thus devoid of dense (hence molecular) gas. In the latest panel, the CMZ has been left almost undisturbed for several tens of megayears. In this phase, collisions tend to dissipate the excess energy of excursions around $x_2$ orbits over $\sim$ tens of Myr, and the gas settles into an $x_2$ ring. This demonstrates that taking into account the bombardment from the bar shocks is essential in understanding the dynamics and structure of the CMZ.

We conclude this section by noting that while all the three mechanisms above certainly play an important dynamical role, they are so intertwined that is difficult to disentangle precisely their individual contributions. The best approach is to compare with an isothermal simulation which effectively switches off the effects of the thermal instability. Overall, the large-scale structures obtained in the current simulations resemble the smooth large-scale structures found in the isothermal case, such as spiral arms and bar shocks, but they are not as smooth and show a considerable amount of substructure, and more unsteady flow. Thus the isothermal simulations appear as a smoothed-out version of the those shown in Fig. \ref{fig:rho2}. 


\subsection{Dynamics in the $(l,v)$ plane}

Figures \ref{fig:lv1} and \ref{fig:lv2} show once again the distribution of atomic hydrogen and CO in the $(x,y)$ plane at $t=181.1\, \Myr$, i.e.\ the same time as in Fig.~\ref{fig:rho2}. They also show how these distributions look when projected into the longitude-velocity plane using the procedure outlined in Section~\ref{synth}. The right-hand panels in both figures are designed to allow one to connect structures in the $(x,y)$ and $(l,v)$ planes by colour-coding different features. Figures \ref{fig:lv1} focuses on the entire bar region, while Fig. \ref{fig:lv2} zooms in on the CMZ itself. It is striking how much smoother the H{\sc i} distribution is compared to the CO, much as we see in real observations of the same two components. Indeed, the bottom panels of Fig. \ref{fig:lv2} show that the CMZ is almost invisible in H{\sc i}, while it is very prominent in CO.

We remind the reader that the procedure used to produce the longitude-velocity projections assumes that the contribution from each {\sc Arepo} cell scales as the mass of the tracer in question divided by the square of the distance, which is equivalent to assuming that the gas is optically thin (and, in the case of CO, that it has a constant excitation temperature). In the case of the H{\sc i} 21~cm line of atomic hydrogen, this is an acceptable approximation, while in the case of CO we expect it to be less accurate (see Section~\ref{synth}). However, here we are not interested in comparing the detailed intensities in the $(l,v)$ diagrams, but only in large-scale features and streams. These are remarkably robust against the choice of the radiative transfer. Figure 6 of \cite{Pettitt++2014} show a longitude-velocity CO diagram from their simulation calculated using two different methods: i) a simple scheme akin to ours ii) a full radiative transfer. The results are strikingly similar. While detailed intensities are different, the overall morphology and large-scale features are identical in the two versions. Thus if we are only interested in the presence and shape of large-scale features and streams it is adequate to use our simple approximation.


\subsubsection{Large-scale structure}

\begin{figure*}
\includegraphics[width=1.0\textwidth]{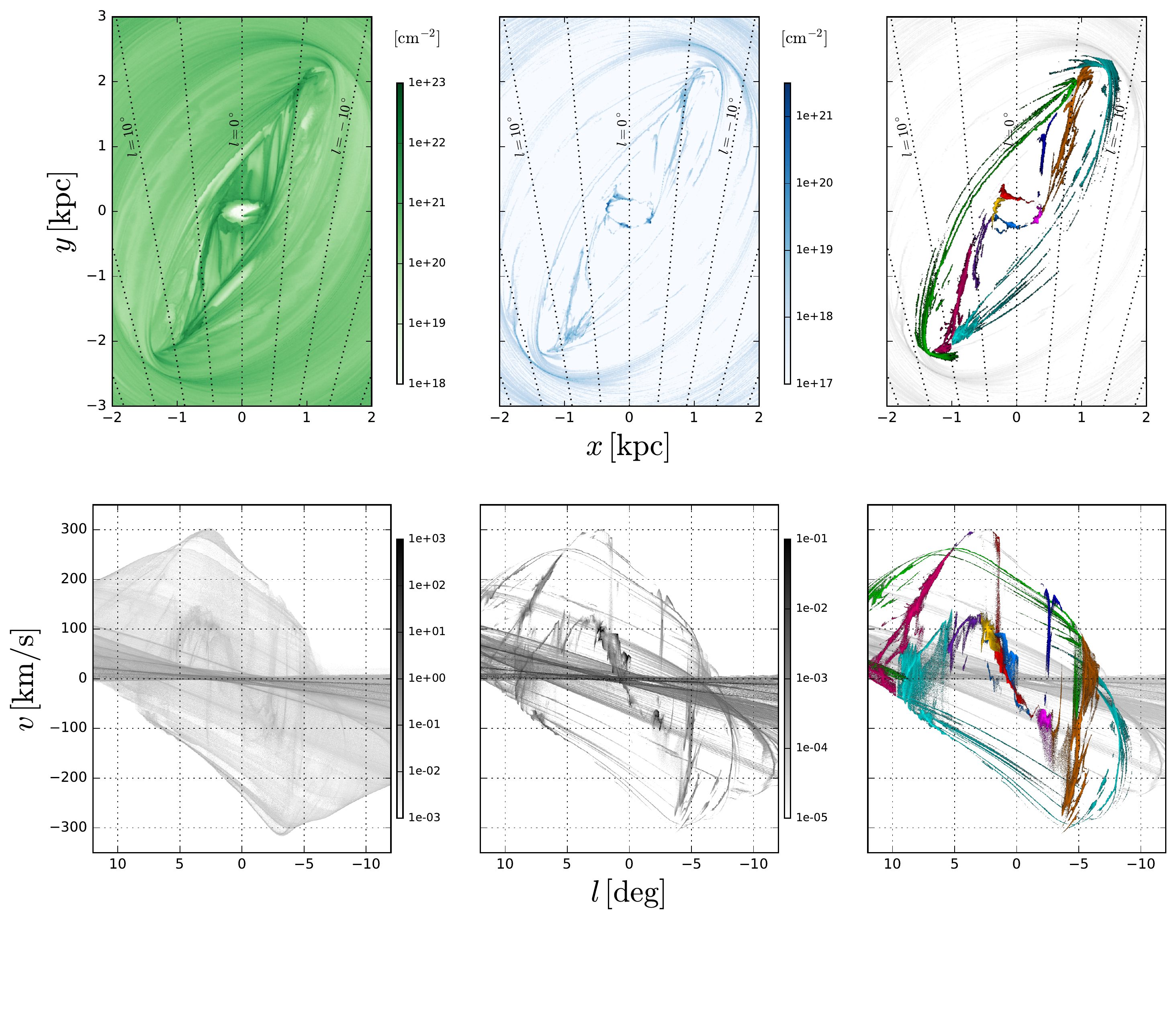}
\caption{The top row shows surface densities for the same snapshot as in Fig. \ref{fig:rho2}. The bottom row shows the corresponding projections to the $(l,v)$ plane. The left panels are for H{\sc i}, the middle panels for CO, and the right panel shows a schematic colouring that helps one to connect features in $(x,y)$ and the $(l,v)$ plane. The Sun is at $(x,y)=(0,-8)\kpc$ in these figures, and is placed such that the bar major axis and the Sun-Galactic Centre line make an angle of $\phi=20\degree$.}
\label{fig:lv1}
\end{figure*}

Several features can be identified in the H{\sc i} and CO $(l,v)$ projections in Fig. \ref{fig:lv1}. A parallelogram whose shape is similar to the one discussed in \cite{Binney++1991} is clearly visible in CO, but reaching out to considerably larger longitudes $l\simeq 9\degree$, $l \simeq -5 \degree$. Two sides of this parallelogram (red and orange in the right panels of Fig. \ref{fig:lv1}) are the projections of the near and far side bar shocks respectively. These shocks would correspond to the dust lanes of the MW bar. Their observational counterparts have been identified in CO, H{\sc i} emission and in dust extinction maps \citep[e.g.][]{Marshall2008}. Their inner portions are also annotated in Fig. \ref{fig:henshaw}. 

Numerous bright ridges running from top-left to bottom-right can be seen in both H{\sc i} and CO in Fig. \ref{fig:lv1}. These are the projections of outer bar-driven spiral arms. Whenever these spiral arms touch the envelope of the emission at $(v>0,l>0)$ and $(v<0,l<0)$ they produce a characteristic ``bump''. Most of these bumps lie at $| l | \geq 12 \degree$ so only the first few of them are visible on the H{\sc i} envelope in these figures. A similar behaviour is found in real observations. These and others features have been discussed in a number of works \cite[e.g.][]{Fux1999,Bissantz++2003,RFC2008,SBM2015c,Li++2016}, and since in this paper we focus on the CMZ, we shall not discuss them further and postpone the study of the larger-scale flow to a future work.

\subsubsection{The CMZ} \label{sec:CMZ}

\begin{figure*}
\includegraphics[width=1.0\textwidth]{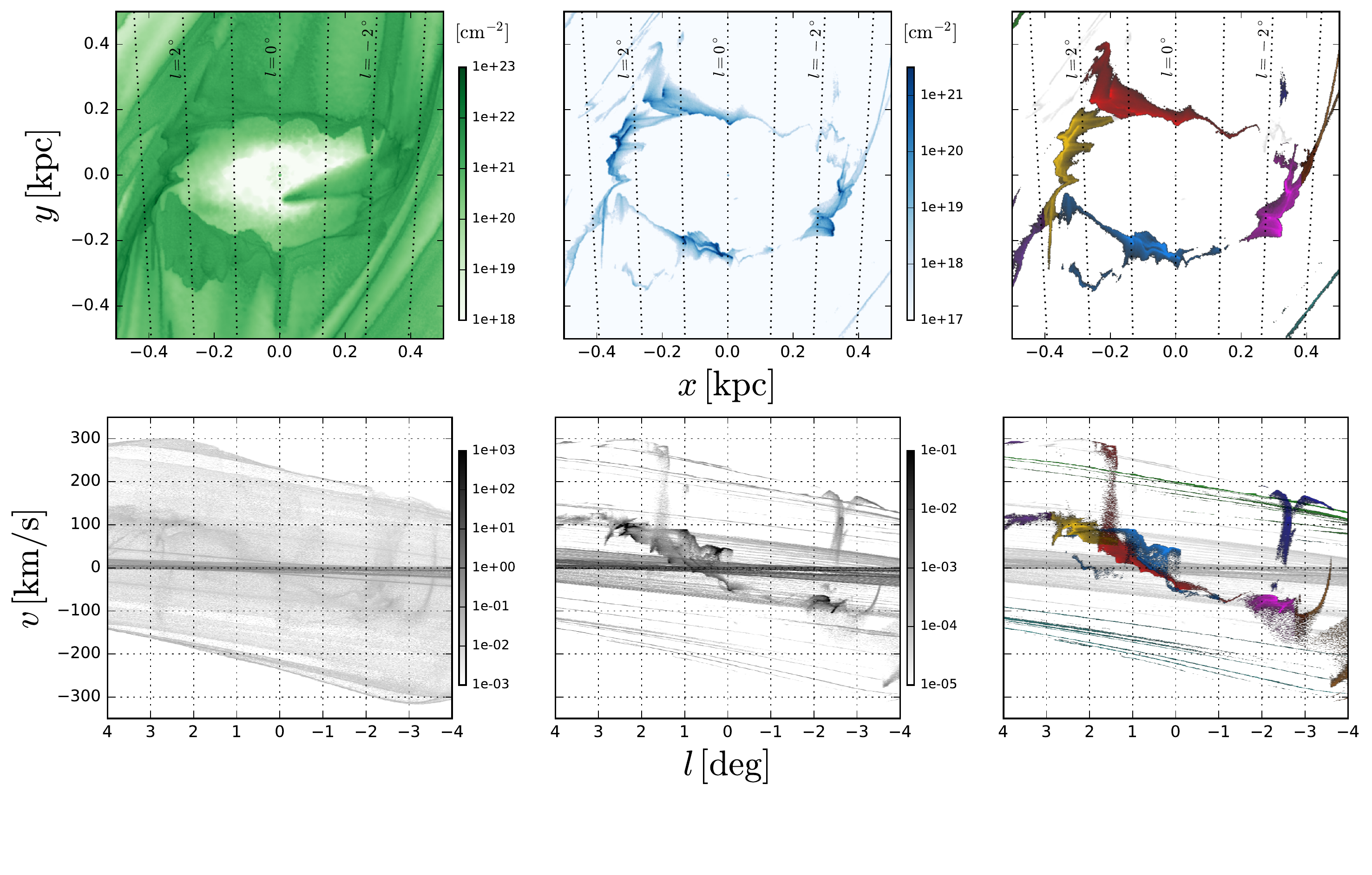}
\caption{The same as in Fig. \ref{fig:lv1}, but zooming into the CMZ.}
\label{fig:lv2}
\end{figure*}

The CMZ $(l,v)$ projection for CO in the middle-bottom panel of Fig. \ref{fig:lv2} is broken up into several independent ``streams'', in remarkable similarity with real observations (compare with the yellow material in Fig. \ref{fig:henshaw}). These streams are the projections of large molecular clouds whose centres of mass approximately move on closed ballistic $x_2$ orbits.

\begin{figure*}
\includegraphics[width=0.5\textwidth]{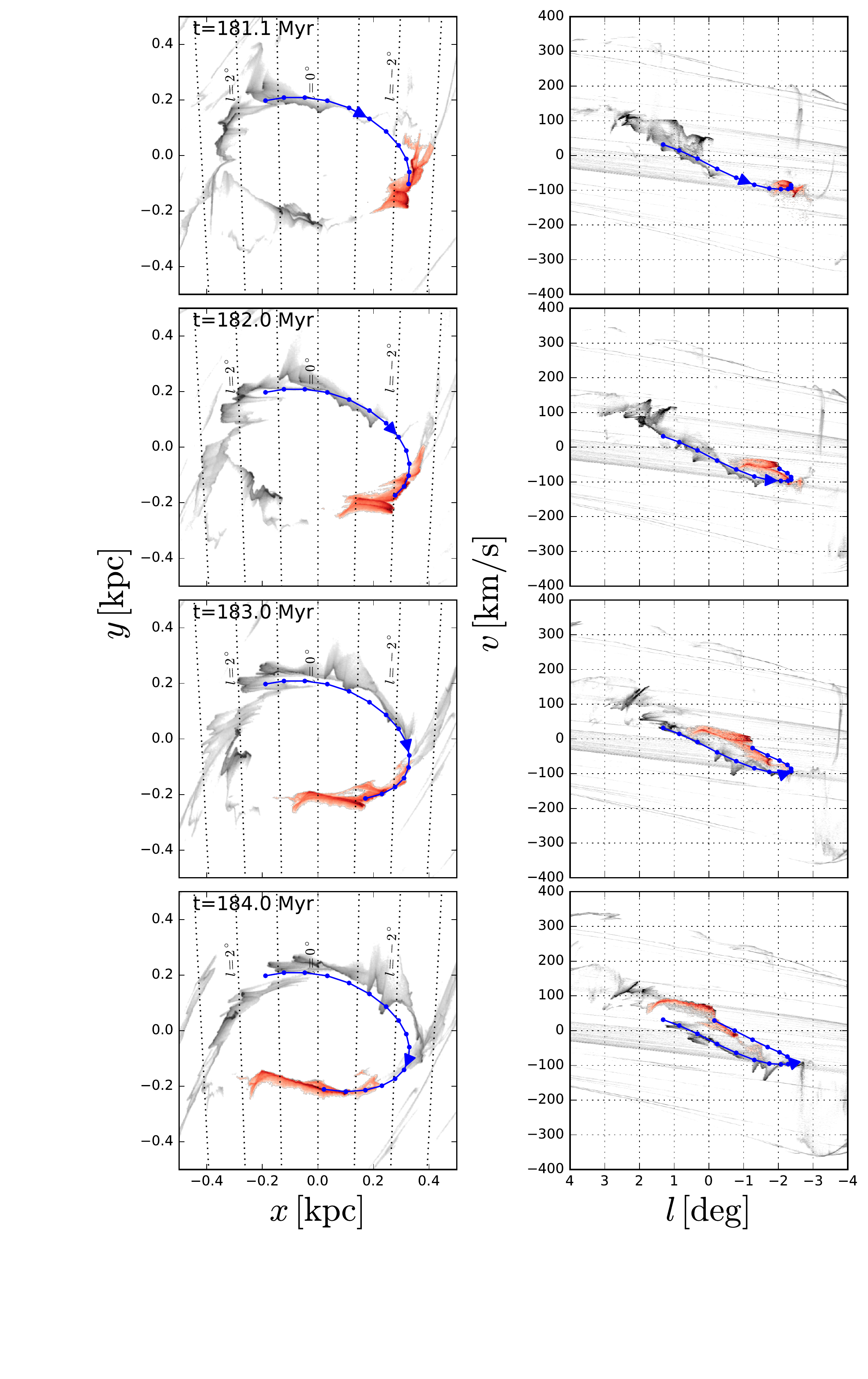}\includegraphics[width=0.5\textwidth]{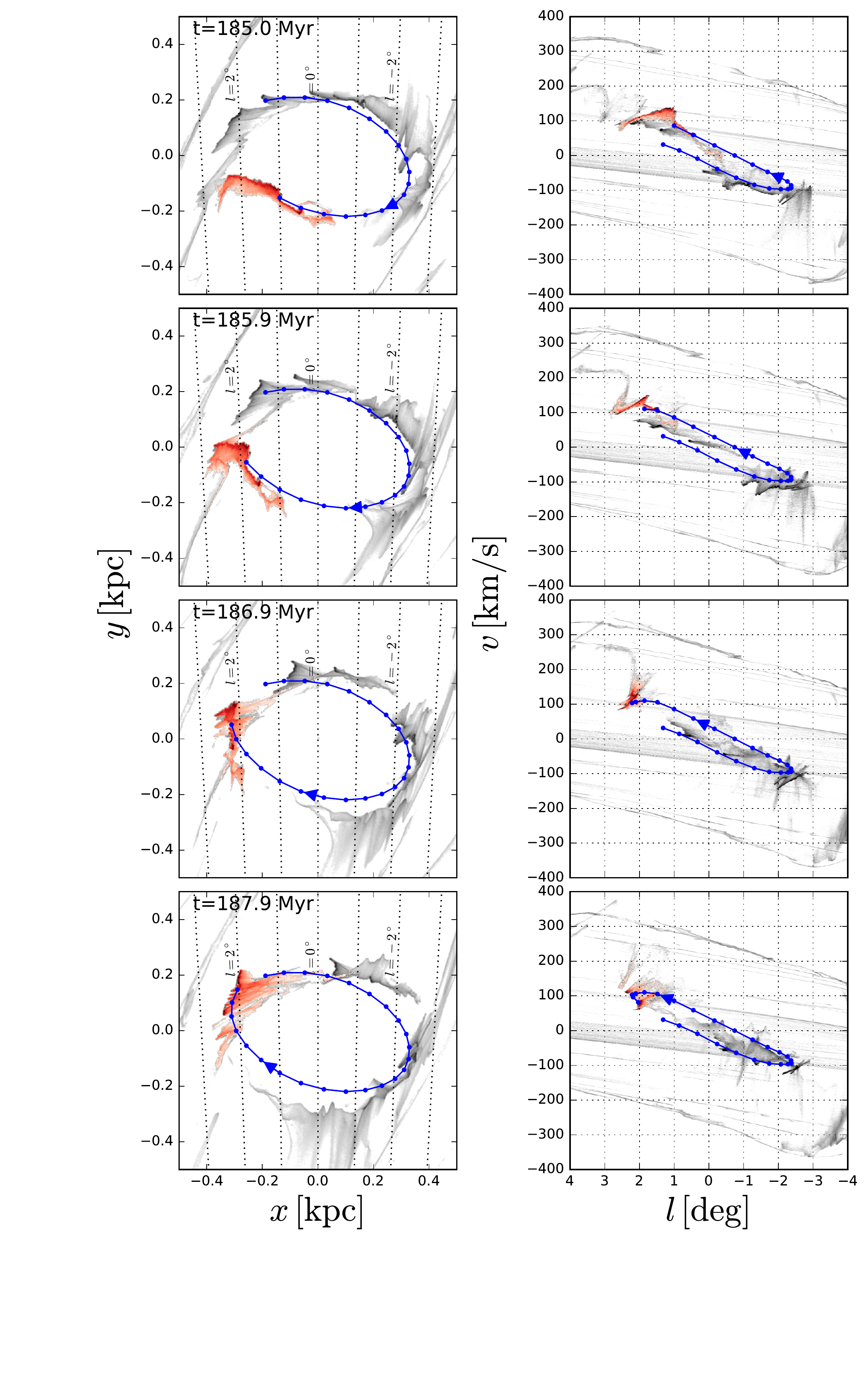}
\caption{A sequence of snapshots of the same simulation shown in Fig. \ref{fig:rho1} separated by approximately $\sim 1 \Myr$. The first and third columns show total surface densities, while the second and fourth columns show CO projections to the $(l,v)$ plane. A molecular cloud is highlighted in red and followed for approximately half a revolution around the centre of the Galaxy. Note that the second panel from the left in the top row corresponds to the same snapshot shown in Figs. \ref{fig:rho2}, \ref{fig:lv1} and \ref{fig:lv2}. The blue line traces the centre of mass of the cloud, which closely follows an $x_2$ orbit, and dots show its position at regular intervals of $0.5 \Myr$.}
\label{fig:lv3}
\end{figure*}

It is evident that the CO distribution is asymmetric both in longitude and velocity in these projections. The appearance of the CMZ is highly transient. This is illustrated in Fig. \ref{fig:lv3} which shows a sequence of snapshots separated by $\sim~1~\Myr$. The snapshot of Fig. \ref{fig:lv2} is included as the first snapshot in the series. A molecular cloud is highlighted in red and followed over half a revolution using tracer particles.

How do features move in the $(l,v)$ plane as a function of time? The red cloud highlighted in Fig. \ref{fig:lv3} shows that there is un underlying anti-clockwise motion in the $(l,v)$ plane.\footnote{Note that these projections are made assuming a constant angle $\phi=20\degree$ between the Sun-Galactic Centre line and the bar major axis. What we would actually observe as the Sun moves on its orbit around the Galactic Centre would be different, as $\phi$ would be changing with time.} However, on top of the basic anti-clockwise motion, the red cloud displays complex additional kinematics due to the excursions around the $x_2$ orbits and spinning motions. A frequently occurring consequence is that ridges in the $(l,v)$ plane \emph{do not} move along the direction in which they are
elongated; thus, they do not always represent a flow parallel to themselves, as one might naively expect. In other words, material does not move \emph{along} a stream. Instead, the stream as a whole \emph{translates} with a component of velocity perpendicular to itself. It can be seen in Fig. \ref{fig:lv3} that indeed streams are often not aligned with the trace of the centre of mass of the parent cloud (blue line).

\begin{figure}
\includegraphics[width=0.5\textwidth]{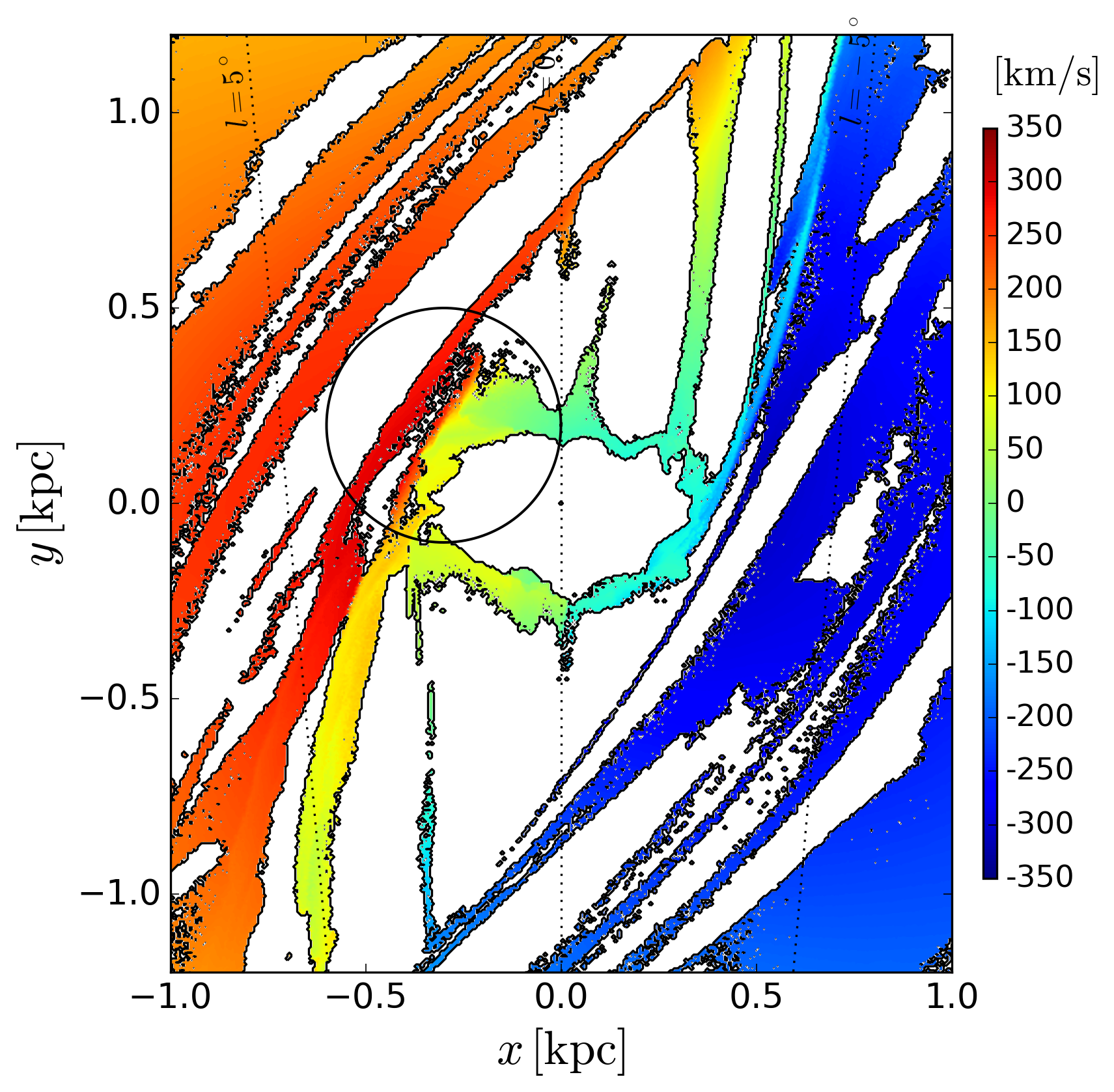}
\caption{Gas coloured by line-of-sight velocity for the same snapshot as in Fig. \ref{fig:rho2}. The snapshot is rotated so that the Sun is located at $(x,y)= (0,-8)\, \kpc$. The black circle denotes the area where the material coming from the near-side shock ``brushes'' the simulated CMZ, causing a transfer of material. Dotted lines indicate the lines of sight for $l=5,0,-5\degree$ respectively.}
\label{fig:losv}
\end{figure}

Figure \ref{fig:losv} shows the molecular gas (H$_2$) coloured by the line of sight velocity as observed from the Sun. The black circle denotes a region where material that has been falling fast down the shock ``brushes'' the CMZ, which is composed of material moving at much smaller velocities. This creates an interface where fast material (red) is in contact with slower material (yellow/green). In this thin layer a transfer of material between fast and slow components takes place, and material falling down the shocks is decelerated. This creates features like the one visible in Fig. \ref{fig:lv2} as the red spur that connects the material at $v\sim 280 \kms$ with the CMZ material at $v \sim 100 \kms$. We therefore interpret the similar features found in the real data in Fig. \ref{fig:henshaw} as ``material brushing the CMZ'' \citep{Fux1999}.

\subsubsection{Proper motions}

\begin{figure}
\includegraphics[width=0.48\textwidth]{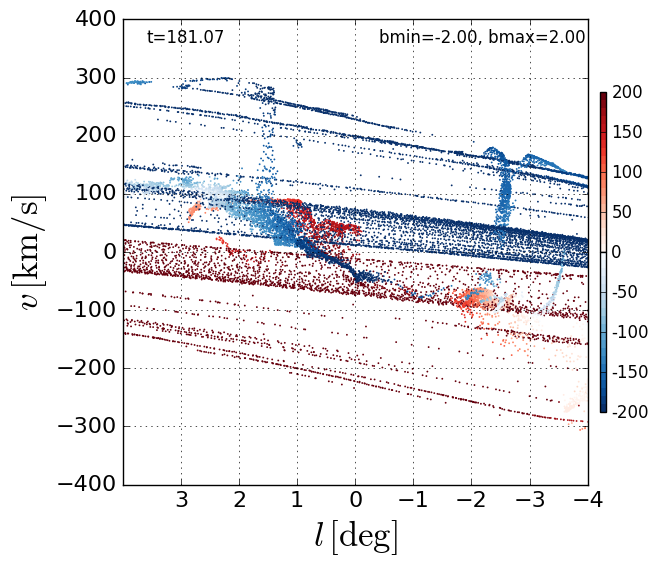}
\caption{Proper motions predictions based on our model. The figure shows the same CO projection of Fig. \ref{fig:lv1}, but coloured by proper velocity. The proper velocity is defined as the component of the velocity perpendicular to the line of sight in the inertial frame of the Galaxy. The colour scale is in $\kms$.}
\label{fig:lv4}
\end{figure}

To discuss motions of features in the $(l,v)$ plane it is also useful to consider proper motions. Figure \ref{fig:lv4} shows predictions for proper motions based on our models. Displayed is the same snapshot as in Fig. \ref{fig:lv2}, but coloured by proper velocity in the $l$ direction. This is defined as the component of velocity perpendicular to the line of sight in the inertial frame of the Galaxy. In Fig. \ref{fig:lv4}, red represents material moving to the left, while blue represents material moving to the right (compare with the motions in Fig. \ref{fig:lv3}). Interestingly, material on $x_2$ orbits (i.e., in the CMZ) rotates on average anti-clockwise, but material on $x_1$ orbits rotates in the opposite sense (i.e., clockwise) in the $(l,v)$ plane. Note that in the $(x,y)$ plane everything rotates in the same direction. It will be interesting to test these predictions against the measured proper motions of masers, which will be available in the upcoming months/years \citep{Ott+2017}.

It is not uncommon for material with opposite proper motions to overlap in the $(l,v)$ plane, especially at the edges of the CMZ near $(l,v) \sim (\pm2\degree,\pm100\kms)$, where material turns around (e.g.\ see overlapping of blue and orange in the bottom-left panel of Fig. \ref{fig:lv2}). Since material moving in opposite directions is usually on opposite sides of the Galactic Centre (e.g.\ Fig. \ref{fig:lv2}), this also means that in the $(l,v)$ plane material on the far side of the Galaxy often lies on top of near-side material. The amount of overlapping is stochastic and depends on how unsteady the flow is, which is significantly affected by the recent ($\sim$ few tens of megayears) bombardment history of the CMZ (see the discussion in Section \ref{sec:bomb}). The overlapping significantly decreases at late times, when bombardment has stopped and the CMZ has settled into a ring. What is the situation in the MW? According to the most likely interpretation of the Connecting Arm as the near side dust lane \citep{Fux1999,Marshall2008} the MW shock lanes are currently filled with gas and therefore bombarding the CMZ (see also Fig. \ref{fig:henshaw}). Hence one must be careful in interpreting the data using kinematical and dynamical models in an attempt to break the near/far degeneracy.

\section{Discussion} \label{sec:discuss}

\subsection{The asymmetry} \label{sec:asymmetry}

It is clear from the projections of Figs. \ref{fig:lv2} and \ref{fig:lv3} that in our models the molecular gas distribution displays a strong degree of asymmetry, both in longitude and velocity. Indeed, as discussed in Section \ref{sec:results}, the molecular gas in the CMZ is broken up into several large molecular clouds, each with a centre of mass that closely follows an $x_2$ orbit, while the overall mass distribution executes additional complex excursions on top of this underlying motion.

\begin{figure}
\includegraphics[width=0.5\textwidth]{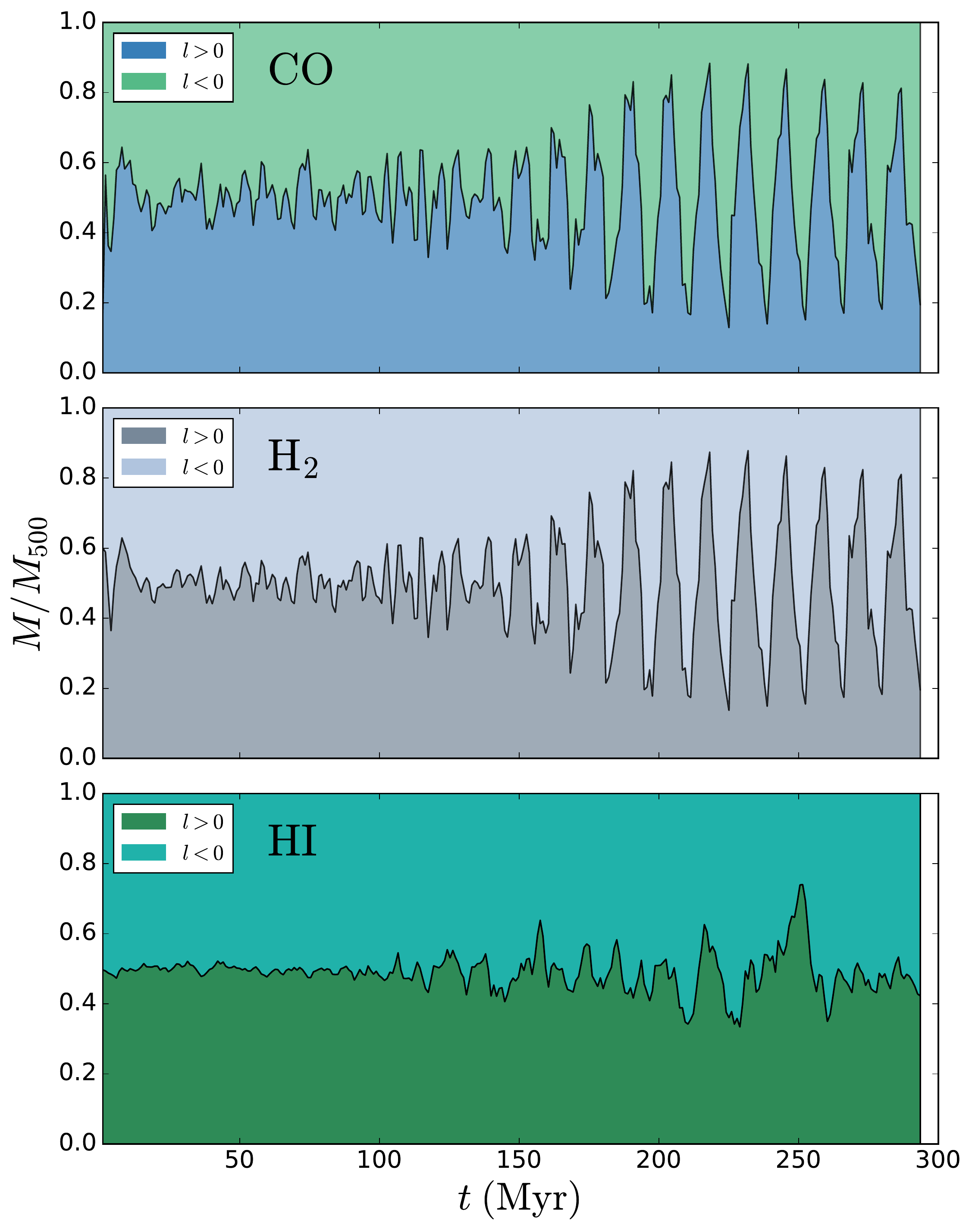} 
\caption{The asymmetry of the CMZ in the variable ISRF simulation as a function of time. $M_{500}$ is defined as the total mass contained at radii $R<500\pc$ in our models. The line separates the fraction of material at positive longitude from the fraction at negative longitudes. From top to bottom: CO, H$_2$, H.
}
\label{fig:asymmetry}
\end{figure}

Figure \ref{fig:asymmetry} quantifies the degree of left-right asymmetry of the simulated CMZ as a function of time. Plotted is the fraction of H, H$_2$ and CO at positive and negative longitudes in the innermost $500\pc$.  It confirms that the distribution of molecular gas is strongly asymmetric. In contrast, the distribution of atomic gas shows only moderate fluctuations. Very similar behaviour is found if one plots the asymmetry in velocity instead of the one in longitude.

The initial density distribution in our models is uniform. However, thermal instability rapidly and stochastically creates dense clouds that quickly form molecular gas. 
This generates moderate fluctuations at $t<100\Myr$. As the bar potential grows stronger, bar shocks form and the gas is further perturbed by the wiggle instability, which creates clumps that crash into the CMZ, disrupting it and creating stronger asymmetries. The conversion of atomic to molecular gas is unsteady, due to density fluctuations that provide widely different amounts of screening against the ISRF. Fluctuations quickly increase in amplitude and become comparable to the observed ones. Between $150 \lesssim t \lesssim 220 \Myr$, the bar potential has been fully switched on, the CMZ is being bombarded by the bar shocks and the degree of asymmetry fluctuates wildly. 

Beyond this time, the supply of fresh gas from the cusped $x_1$ orbit decreases and the bombardment slows down (see Section \ref{sec:bomb}). The CMZ molecular clouds then dissipate the excursions around $x_2$ orbits and settle into an $x_2$ ring. Now the asymmetry behaves periodically, with a period of approximately $T\simeq 15 \Myr$. This simply corresponds to the orbital period of clouds on $x_2$ orbits. Note that since Fig. \ref{fig:asymmetry} assumes a constant angle between the Sun-GC line and the bar major axis this is \emph{not} the period that we would observe by following the Sun along its orbit and pointing our telescopes towards the CMZ. The latter would be slightly larger by approximately $\sim 1 \Myr$ since the Sun partly compensates for the rotation of the CMZ by rotating around the Galactic Centre more slowly but in the same direction. 

Thus, there are two characteristic timescales associated with the asymmetry: the orbital period of $x_2$ orbits ($T\simeq 15 \Myr$) and the timescale associated with the bar shock feeding ($T \sim 1\mhyphen 5 \Myr$).

It is remarkable that fluctuations so large are obtained despite the initial conditions and the bar potential being point-symmetric with respect the Galactic Centre, except from some moderate random noise. The entire asymmetry is therefore the result of unsteady flow, triggered by inhomogeneities due to the thermal instability and enhanced by hydrodynamical instabilities and unsteady conversion of material from atomic to molecular form.

We have performed further tests to confirm our interpretation. To test whether it is really inevitable to reach such a degree of asymmetry, we have restarted the ``variable'' ISRF simulations from $t=300\Myr$ by keeping the same velocity field but resetting densities, temperature and chemical abundances everywhere so that their distributions are completely uniform. This also avoids the gradual introduction of the bar potential. We have found that strong asymmetries again develop during the first few tens of megayears. This confirms that the ``symmetric'' state is not a stable and durable one.

We argued above that thermal and hydrodynamical instabilities are both important in developing the asymmetry. To test this, we have run a simulation in which the initial gas density vanishes outside $R=500\pc$. This setup eliminates CMZ feeding from the bar shocks while keeping the thermal instability present in the system. In this case, as one would expect, the CMZ simply settles into a smooth ring similar to that obtained at late times in Fig. \ref{fig:rho1}, and with a much less skewed mass distribution that reaches a left/right ratio of $\sim 6/4$, significantly lower than  in the simulations discussed above and in observations of the real CMZ. This demonstrates that the thermal instability alone is not sufficient to generate the observed degree of asymmetry.

\begin{figure}
\includegraphics[width=0.48\textwidth]{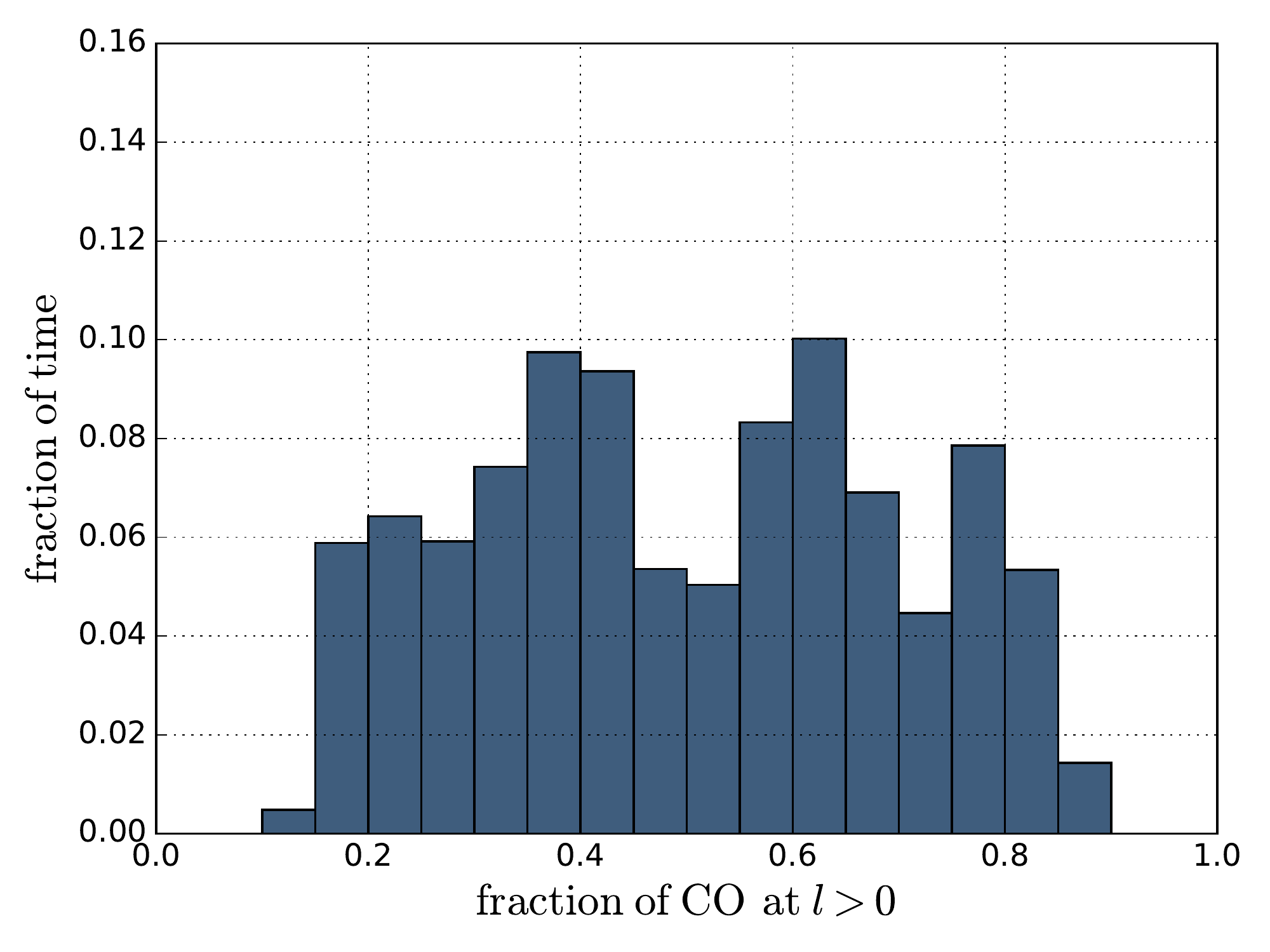} 
\caption{Histogram showing the fraction of time during which a given degree of asymmetry is found. On the $x$ axis, we show the fraction of CO within $R<500\pc$ which lies at positive longitudes, normalised to the total CO mass within the same region. On the $y$ axis, we show the fraction of time spent in each bin. For example, the fraction of time during which between 70 and 75$\%$ of the CO lies at positive longitude is $\sim10\%$. The plot is made using results from the variable ISRF simulation and considers only snapshots at $t>150\, \Myr$, i.e.\ the period after the bar potential has been fully switched on.}
\label{fig:asymmetry2}
\end{figure}

Finally, Fig. \ref{fig:asymmetry2} shows the fraction of time during which a given range of asymmetry is attained in our models. It suggests that for large fractions of the time asymmetries are expected, and that an asymmetric CMZ is not a peculiar but rather a typical state of our Galaxy. An asymmetry as large as the one presently observed in our Galaxy is not particularly uncommon.

\subsection{The morphology of the CMZ}

In contrast to the smooth, coherent nuclear spiral arms present in the simulations of \cite{Ridley+2017}, the appearance of the CMZ in molecular gas in our simulations is patchy and ring-like. The nuclear spiral arms of \cite{Ridley+2017} are the result of constructive interference between librations of the gas around the $x_2$ orbits \citep{SBM2015b}, and therefore require pressure support to form. Our simulations lack pressure support in molecular gas because we lack an important ingredient, star formation. By injecting small-scale turbulence into the ISM, which is absent in our simulations (see also Sect. \ref{sec:fails}), stellar feedback increases the pressure support of the molecular component, favouring the development of molecular nuclear spirals. It may also provide a small amount of turbulent viscosity.

In external galaxies the morphologies of nuclear discs are complex, often showing both ring-like and spiral-like structures with a strong dependence on the tracer used \cite[eg.][]{Izumi+2013}. In practice it is likely that the CMZ of the MW displays aspects of both morphologies, and given the lack of a precise definitions it is far more useful to consider the dynamics of the CMZ in the context of the larger scale gas flow and the processes that govern it.

Other morphologies for the CMZ have been proposed, such as an open orbit \citep{Kruijssen+2015}. We have found no evidence of massive clouds following open orbits after entering the CMZ. Their centres of mass follow closed $x_2$ orbits, on top of which excursions take place leading to complex $(l,v)$ diagrams (c.f. Fig. \ref{fig:lv3}). However it may be that the addition of stellar feedback and/or magnetic fields would help to lift material up out of the plane, leading to more open orbits.

\subsection{Runs with different strengths of the interstellar radiation field} \label{sec:lvh}
\begin{figure*}
\includegraphics[width=1.0\textwidth]{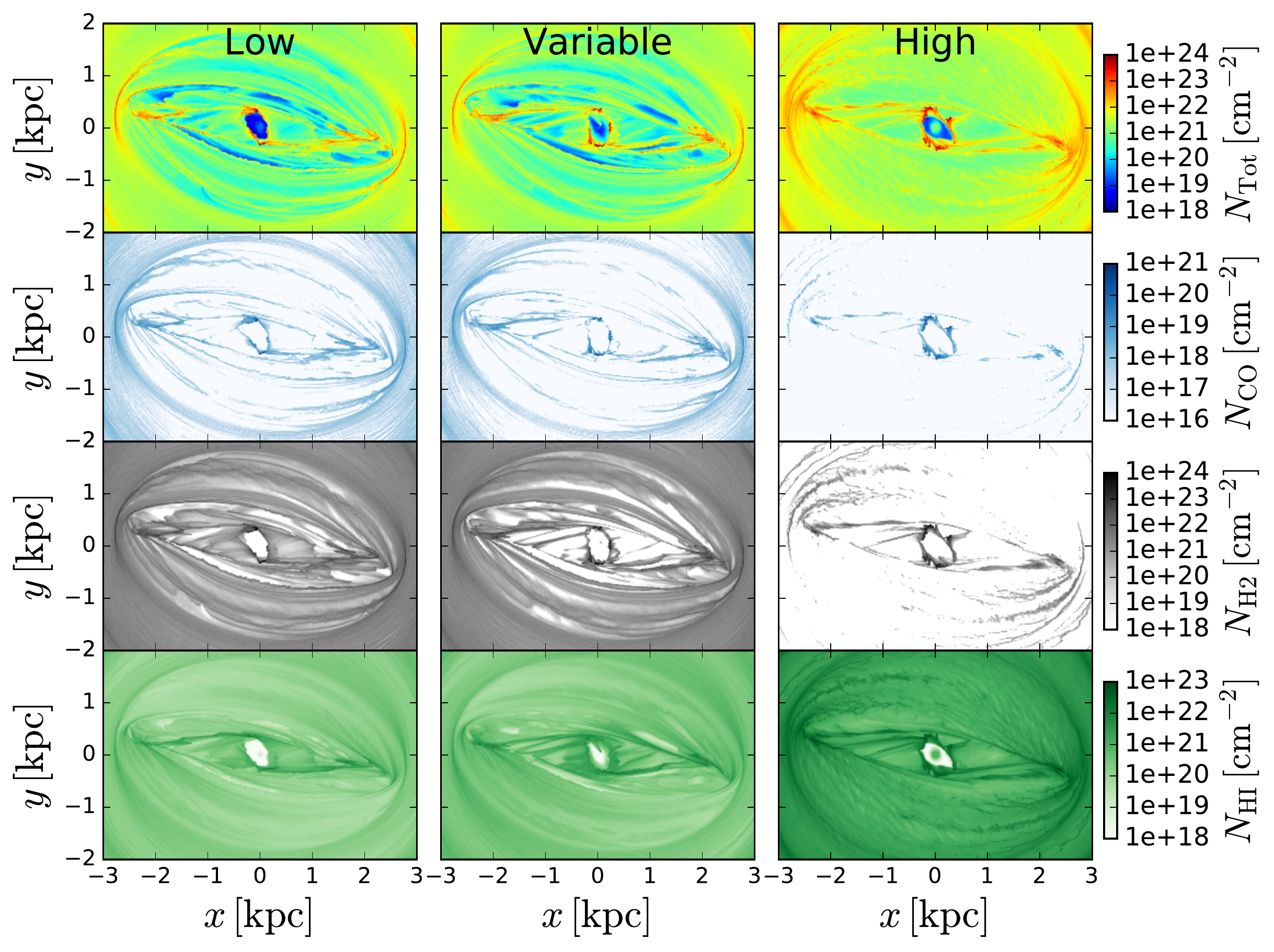}
\caption{Comparison at $t=181\Myr$ of the three simulations presented in the paper, which differ from the imposed ISRF. From left to right, low, variable and high ISRF (cf Fig. \ref{fig:ISRF}).}
\label{fig:compare_lvh}
\end{figure*}

So far, we have only discussed the ``variable'' simulation. The main difference between this and the ``low''  (``high'') simulation is that there is more (less) molecular gas. Indeed, in practice the strength of the externally imposed ISRF simply controls the amount of molecular gas in our simulations. This can be seen from Figure \ref{fig:compare_lvh}, which compares the ``low'', ``variable'' and ``high'' simulations.

The dynamical effects that we have discussed in this paper are largely insensitive to the value of the ISRF, to the point that the two limiting cases in Fig. \ref{fig:ISRF} produce a CMZ with comparable degrees of asymmetries. Thus the discussion and conclusions presented in this paper using the ``variable'' simulations are unaffected by considering the ``low'' or ``high'' simulations. 

\subsection{Implications for external galaxies}
Our simulations produce results that vary strongly with time.  This is particularly pronounced in the central regions, where the detailed predictions of the simulations changes significantly from snapshot to snapshot. This suggests that the same may happen in external galaxies. Here we have employed an external potential that is tuned for the Milky Way, but the fact that an asymmetry develops is not particularly sensitive to the details of the external potential
employed. Hence, we would expect a significant fraction of external barred galaxies to show a significant degree of asymmetry in the molecular gas distribution of their central regions. Indeed, some of them (e.g. M83) show quite irregular nuclear regions \citep[e.g.][]{Lundgren+2004,Frick+2016}. In M83 it may be that the nuclear region has been recently swept by a big molecular mass that has been falling along
the dust lanes. Others (e.g.\ NGC1097) appear more regular, but a closer look with ALMA \cite[e.g.][]{Izumi+2013,Fathi+2013} reveals that the gas distribution displays more substructure than is apparent  at first glance.

\subsection{Shortcomings of our model} \label{sec:fails}

\subsubsection{Vertical structure} \label{sec:vertical}

\begin{figure*}
\includegraphics[width=0.85\textwidth]{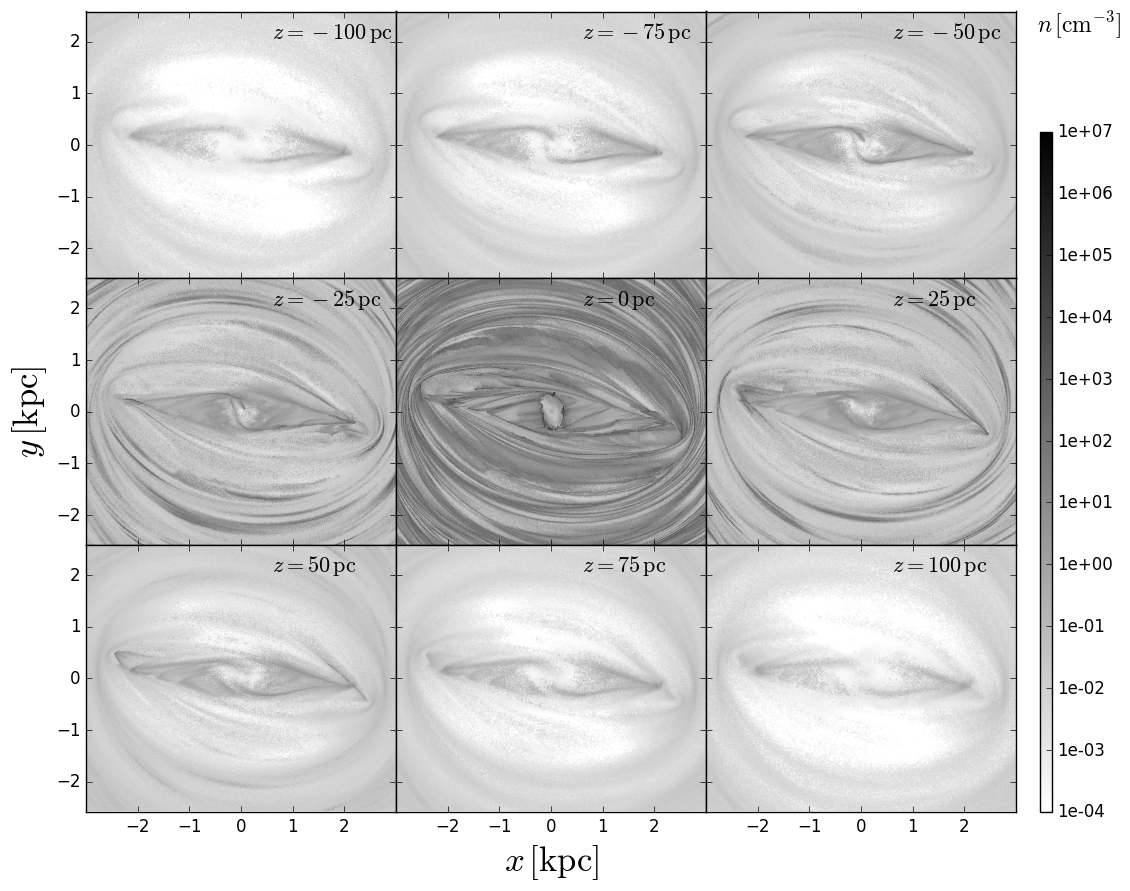}
\caption{Slices of total gas density along planes parallel to the plane $z=0$ for the same snapshot of Fig. \ref{fig:rho2}.}
\label{fig:rhoslices}
\end{figure*}

While the dynamics in the $(x,y)$ plane is very rich in our models, in the vertical direction not much happens. Figure \ref{fig:rhoslices} shows slices of the total density at different heights. It shows that the dense molecular component is much more confined to the Galactic plane than the more diffuse component. The more diffuse component has much in common with the appearance of the isothermal simulations in \cite{Ridley+2017}. Thus, the addition of the third dimension by itself does not seem to produce major differences in the gas flow. This confirms that treating the gas as a razor-thin disc is a fairly good approximation if we are only interested in motions within the Galactic plane.

In our simulations, the thickness of the gaseous layer is $\sim \rm 15\pc$ for the molecular gas and $\sim300\pc$ for the atomic gas. These are smaller than the observed counterparts. In particular, the observed thickness of dense CO is $\sim 90\pc$ \citep[][]{HeyerDame2015}. Our inability to reproduce the correct molecular gas scale height is unsurprising, since previous simulations of the
Galactic disc including similar physics \citep[e.g.][]{Dobbs+2011} also yield a gas distribution that is much thinner than the observed one. The missing ingredient is most likely an appropriate level of small-scale turbulence driven by stellar feedback -- in the absence of this, there is insufficient pressure  to prevent the gas from collapsing into a thin layer. The same mechanism is likely to also be responsible for the \cite{Larson1981} size-linewidth relation, which is also not reproduced in our models. We will be able to test this in future simulations that include star formation and stellar feedback, but these are beyond the scope of our current study. 

Note however that the H{\sc I} layer is significantly thinner in the inner parts ($R \lesssim 2.4\kpc$) of the Galaxy \citep{LockmanMcClure2016}, and the walls that surround this thin layer trace the walls of the Fermi Bubbles \citep{BHC2003,Ackermann+2014}. It may be that this thinness is explained by a very low amount of star formation, which we would expect in this region in analogy with barred external galaxies. The fact that the Fermi bubbles trace the walls of this region might be due to the fact that the Fermi Bubbles simply expand to fill the void previously created by the lack of star formation.

Projected $(l,v)$ streams are more coherent and sharper in our models than in real observations. Our neglect of the thermal velocity dispersion of the gas, which would essentially result in a small gaussian smoothing of the final $(l,v)$ distributions, marginally contributes to this sharpness. But the sharpness is mostly attributable to the simulations not including small-scale turbulence. Stellar feedback maintains random velocities of a few km/s \citep{Shetty12}. Its absence from the simulations makes the simulated molecular layer too thin in the direction perpendicular to the Galactic plane, which results in giant molecular clouds in our simulations that are more coherent than those in the real Galaxy.

\subsubsection{Size of the CMZ}

The reader will have noticed that the CMZ in our models is too large. It extends to $ | l | \approx 2.5\degree$, while in the real Galaxy it is closer to $ | l | \approx 1.5\degree$. The size of the CMZ in our models is controlled mainly by the mass of the bulge and the properties of the bar, such as its pattern speed. Thus this problem is likely to be fixable by fine-tuning the parameters of the large scale potential. However, to provide a more direct comparison with the work of \cite{Ridley+2017} and since our simulations are computationally expensive to run, we have not fine-tuned this parameter to reproduce the CMZ size more accurately. This is probably best done with simpler simulations than those presented in this work, and once a better potential is found it can be used to run a more computationally expensive simulation such as the one presented here.

\subsubsection{Mass of the CMZ}

Due to the lack of star formation, in our simulations the gas simply accumulates in the CMZ, where it sits forever. In the real Galaxy some of this gas would turn into stars, but since star formation is not included in our simulations we end up overestimating the total amount of molecular gas in CMZ. The interstellar radiation field dissociates part of this gas, but this effect is too weak to compensate for the absence of star formation.

\section{Conclusion}
\label{sec:conc}

We have shown that the observed asymmetry in the molecular gas distribution in the Central Molecular Zone of the Milky Way can be neatly explained by unsteady flow originating from a combination of 
\begin{enumerate}
\item A corrugation instability of shock fronts known as the wiggle instability.
\item The thermal instability.
\item Bombardment of the CMZ from the bar shocks. 
\end{enumerate}
These effects work in conjunction with unsteady conversion of atomic to molecular gas.

We have used high-resolution 3D hydrodynamical simulations coupled to a time-dependent chemical network. We found that, despite our initial conditions and the bar potential being point symmetric with respect to the Galactic Centre, the molecular gas distribution spontaneously becomes asymmetric. The observed unsteady flow requires both thermal and wiggle instabilities in order to produce fluctuations large enough to explain the observations. The thermal instability arises in our model because our cooling function effectively yields a two-phase medium of the type envisaged by \cite{Field+1969}. Hydrodynamical instabilities occur because bar shocks are formally unstable \citep[e.g.][]{KimKimKim2014,SSSTK}. Due to this instability, perturbations are amplified when crossing the bar shocks. Once the flow becomes unsteady, gas is stochastically converted from atomic to molecular form, producing quantities of molecular gas that fluctuate wildly. This also naturally predicts the widespread presence of shock tracers observed in the CMZ \citep[][]{Jones+2012}.

In the simulations, fluctuations of amplitude comparable to the observed asymmetries occur for a large fraction of the time, and suggest that the present time is not a peculiar but rather a typical moment in the life of our Galaxy.

Our results are remarkably robust against uncertainties in the strength of the ISRF, and the asymmetry is essentially the same in the ``low'' and ``high'' simulations which correspond to extreme limiting values of the ISRF. 

The presence of stellar feedback may work alongside the mechanism proposed here to create further unsteady flow. However, we have shown that even in the absence of stellar feedback, the asymmetry appears to be unavoidable.

Despite our models being three-dimensional, they fail to explain the vertical structure of the observed ISM. The gas layer is too thin because the simulations lack small-scale turbulence, probably because of the neglect of stellar feedback, which is also the likely origin of \cite{Larson1981} relation. Also the tilt of the inner ($R\lesssim 3\kpc$) molecular gas layer \citep[e.g.][]{lisztburton1980} remains unexplained.

Another shortcoming of our models is that the CMZ disc is larger that the observations imply. This problem can most likely be cured by carefully tuning the stellar potential. However, that is best done with simpler simulations that are less computationally expensive than the ones presented in this paper.

Directions of future work include: 
\begin{enumerate}
\item Tuning the stellar potential to produce a $x_2$ disc that has the same size as the observed CMZ. Once this is done, by comparing molecular clouds formed in the simulation with prominent molecular clouds seen in observations a de-projection can be attempted to create a face-on picture of the CMZ.
\item Adding further physical ingredients to determine whether the vertical structure of the CMZ can be explained. This includes investigating triaxial bars whose axis may not lie in the plane of the Galaxy, to see whether this could explain the observed tilt \citep{OstrikerBinney1989}.
\item Adding sink particles to study where and how star formation takes place in a gas flow in a barred potential. Intermittent feeding from the bar shocks could result in episodic star formation \citep[see also][]{KrumholzKruijssen2015}.
\item Studying the accretion onto the black hole Sgr A*, which given the general characteristics of the gas flow is likely to be intermittent \citep[e.g.][]{KingPringle2006}. This can be done for example by adding the potential due to the black hole Sgr A*, which was unimportant on the scales of interest in this paper, and measuring the mass accreted.
\end{enumerate}

Movies of our simulations in the $(x,y)$ and $(l,v)$ planes are available at the following link: \url{http://www.ita.uni-heidelberg.de/~mattia/download.html}

\section*{Acknowledgements}

The authors thank Cara Battersby, Paul Clark, Tom Dame, Christoph Federrath, Adam Ginsburg, Jonny Henshaw, Jens Kauffmann, Diederik Kruijssen, Steve Longmore, Naomi McClure-Griffiths, Thushara Pillai, Steve Shore for insightful comments and discussions, the referee Nick Gnedin for valuable suggestions that improved the paper, Volker Springel for allowing use of the code {\sc Arepo} and Jonny Henshaw and Steve Longmore for kindly providing the NH$_3$ data. MCS, RGT, SCOG, and RSK acknowledge support from the Deutsche Forschungsgemeinschaft via the Collaborative Research Centre (SFB 881)
``The Milky Way System'' (sub-projects B1, B2, and B8) and the Priority Program SPP 1573 ``Physics of the Interstellar Medium'' (grant numbers KL 1358/18.1, KL 1358/19.2, and GL 668/2-1). RSK furthermore thanks the European Research Council for funding in  the ERC Advanced Grant STARLIGHT (project number 339177). MR and JM acknowledge support from ERC grant number 321067, while JJB 
is supported by the ERC under the European 
Union's Seventh Framework Programme (FP7/2007-2013)/ERC grant agreement no.~321067. The authors acknowledge support by the state of Baden-W\"urttemberg through bwHPC and the German Research Foundation (DFG) through grant INST 35/1134-1 FUGG.

\def\aap{A\&A}\def\aj{AJ}\def\apj{ApJ}\def\mnras{MNRAS}\def\araa{ARA\&A}\def\aapr{Astronomy \&
 Astrophysics Review}\def\apjs{ApJS}\def\apjl{ApJ}\def\pasj{PASJ}\def\nat{Nature}\def\prd{Phys. Rev. D}
\def\ssr{Space Sci. Rev.}\def\pasp{PASP}\def\aaps{A\&A Supplement series}
\bibliographystyle{mn2e}
\bibliography{bibliography}

\appendix

\section{Some numerical experiments} \label{appendix:a}

\begin{figure*}
\includegraphics[width=1.0\textwidth]{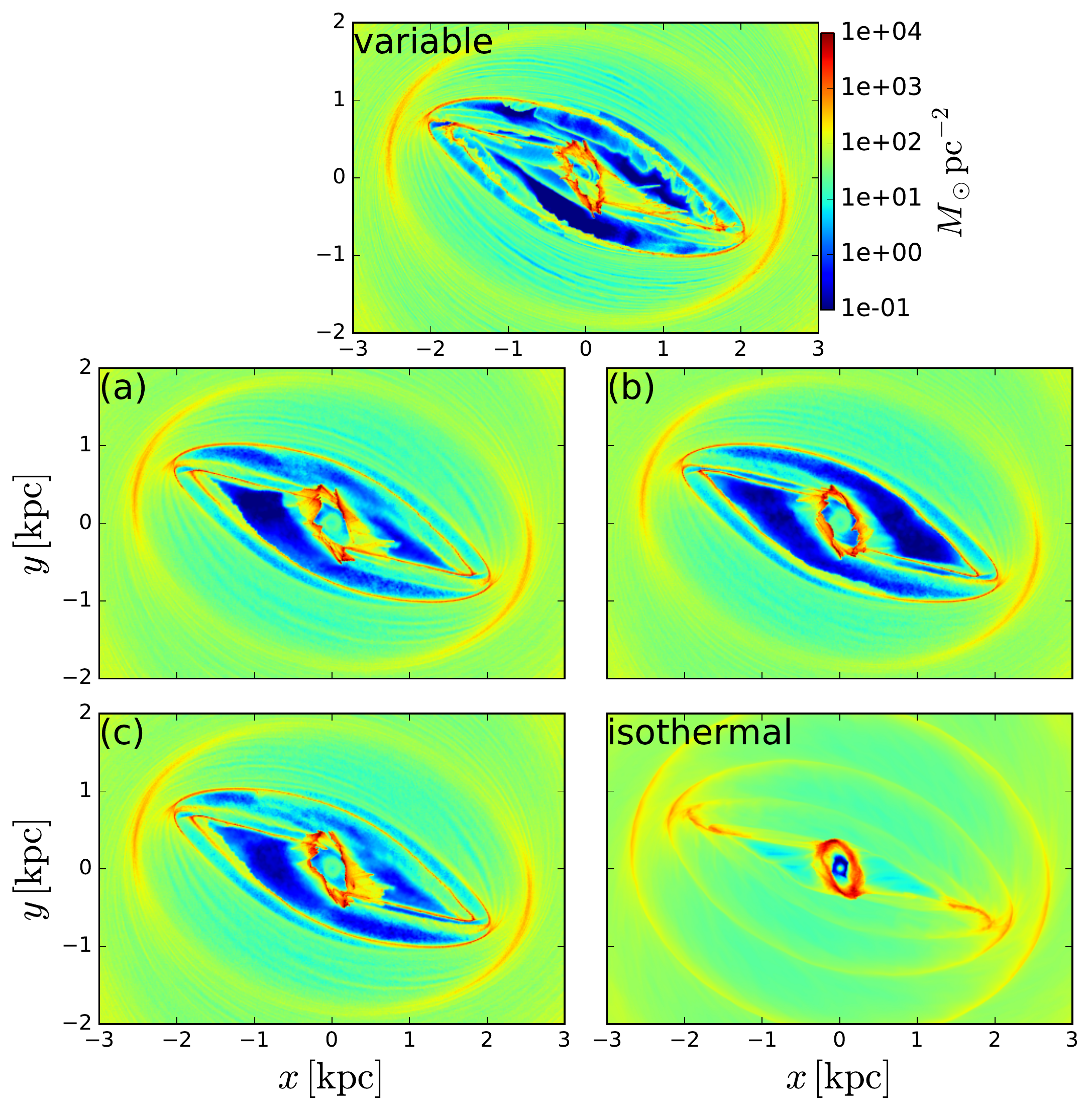}
\caption{Total surface density of various simulations at $t=132 \Myr$. {\bf (variable)} the simulation discussed in the main text. {\bf (a)} Same as (variable), but at a lower resolution. {\bf (b)} Same as (a), but for an initial density distribution which is exactly uniform rather than containing small fluctuations due to Poisson sampling. {\bf (c)} Same as (a), but for an initial surface density distribution that radially declines exponentially after $R = 4\kpc$ with a scalelength $R=7\kpc$, similar to the observed value for the MW. {\bf(isothermal)} Same as (a), but for an isothermal equation of state with $\cs=10\kms$ and disabling the chemistry.}
\label{fig:appendix1}
\end{figure*}

Figure \ref{fig:appendix1} compares the surface density distribution of various simulations at the same time $t=132 \Myr$. The top panel shows the ``variable'' simulation that is discussed in the main text. Panel (a) shows the same simulation at 10 times lower resolution: the target mass is $1000 \, \rm M_\odot$ per cell instead of $100 \, \rm M_\odot$ per cell. At this lower resolution, the bar shocks and the spiral arms starting at and extending out from the tips of the shocks have a smooth appearance, in contrast to the much more zig-zagged appearance the same features have at higher resolution in the top panel. The most probable reason is that the wiggle instability on these features cannot be resolved at the lower resolution (The wiggle instability is very sensitive to resolution, see e.g. \citealt{SBM2015a}). Both simulations appear to be able to resolve the unsteady flow in the CMZ, where the wiggle instability is stronger; however its precise role is difficult to disentangle from the other effects discussed in Section \ref{sec:bomb}. 

Panel (b) shows the same simulation as (a) but for initial conditions in which the density distribution is \emph{exactly} uniform. In other words, we have removed the small random noise caused by the way initial conditions were generated as described in Section \ref{sec:ic}. Panel (c) shows the same simulation as in (a), but for an initial surface density which exponentially declines after $R=4 \, \kpc$ with a scalelength $R=7 \, \kpc$, compatible with the observed decline of the H{\sc i} surface density in the MW. The results of (b) and (c) are essentially indistinguishable from the results of (a), in particular for what concerns the unsteady dynamics of the CMZ, proving that i) the initial random noise does not affect the final degree of asymmetry ii) it is not a radial redistribution of gas from the outer parts of the Galaxy to the inner parts which creates the effects discussed in the main text, as one would expect since it is known that the corotation resonance and the 4:1 resonance work as ``barriers'' and prevent gas from the outer parts from reaching the regions at $R \sim 3-4 \kpc$ \citep[e.g.][]{Sellwood2013}.

Finally, the panel (isothermal) shows the same as (a), but imposing an isothermal equation of state with $\cs = 10 \kms$ and switching off the chemistry. This time, the CMZ shows only mild signs of unsteady flow and becomes only weakly asymmetric. Part of the reason is that at this lower resolution the wiggle instability is not well resolved, and indeed we have verified that at higher resolution the amount of unsteadiness significantly increases. The other reason is that the thermal instability cannot enhance the wiggle instability in the way discussed in Section \ref{sec:bomb}.

%
%

\end{document}